\documentclass[preprint,12pt]{elsarticle}
\usepackage{amsthm}
\usepackage{amsmath}
\allowdisplaybreaks[4]
\usepackage{svg}
\usepackage{booktabs}
\usepackage{threeparttable}
\usepackage{tablefootnote}
\usepackage{url}
\usepackage{multirow}
\usepackage[algoruled,linesnumbered]{algorithm2e}
\usepackage{cite}
\usepackage{tikz}
\newcommand*{\circled}[1]{\lower.85ex\hbox{\tikz\draw (0pt, 0pt)%
    circle (.47em) node {\makebox[0em][c]{\small #1}};}}

\newtheorem{definition}{Definition}
\newtheorem{theorem}{Theorem}

\usepackage{amssymb}
\usepackage{xcolor}
\newcommand{\revised}{\textcolor{black}}
\newcommand{\newrevised}{\textcolor{black}}

\journal{Artificial Intelligence}

\begin{document}

\begin{frontmatter}



\title{Incremental Measurement of Structural Entropy for Dynamic Graphs}


\author[1]{Runze Yang}
\author[2]{Hao Peng\corref{c1}}\cortext[c1]{Corresponding author}\ead{penghao@buaa.edu.cn}
\author[3]{Chunyang Liu}
\author[1]{Angsheng Li}
\affiliation[1]{organization={State Key Laboratory of Software Development Environment, School of Computer Science and Engineering, Beihang University},
            city={Beijing},
            postcode={100191}, 
            country={China}}

\affiliation[2]{organization={School of Cyber Science and Technology, Beihang University},
            city={Beijing},
            postcode={100191}, 
            country={China}}

\affiliation[3]{organization={Didi Chuxing},
            city={Beijing},
            postcode={100193}, 
            country={China}}

\begin{abstract}
Structural entropy is a metric that measures the amount of information embedded in graph structure data under a strategy of hierarchical abstracting. 
To measure the structural entropy of a dynamic graph, we need to decode the optimal encoding tree corresponding to the best community partitioning for each snapshot. 
However, the current methods do not support dynamic encoding tree updating and incremental structural entropy computation. 
To address this issue, we propose \textit{Incre-2dSE}, a novel incremental measurement framework that dynamically adjusts the community partitioning and efficiently computes the updated structural entropy for each updated graph. 
Specifically, \textit{Incre-2dSE} includes incremental algorithms based on two dynamic adjustment strategies for two-dimensional encoding trees, i.e., \textit{the naive adjustment strategy} and \textit{the node-shifting adjustment strategy}, which support theoretical analysis of updated structural entropy and incrementally optimize community partitioning towards a lower structural entropy.
We conduct extensive experiments on $3$ artificial datasets generated by \textit{Hawkes Process} and $3$ real-world datasets. 
Experimental results confirm that our incremental algorithms effectively capture the dynamic evolution of the communities, reduce time consumption, and provide great interpretability.

\end{abstract}


\end{frontmatter}


\section{Introduction}\label{sec:intro}
In 1953, Shannon~\citep{shannon1953lattice} proposed the problem of structural information quantification to analyze communication systems.
Since then, many information metrics of graph structures~\citep{rashevsky1955life-c1, trucco1956note-c2, mowshowitz1967entropy, korner1973coding-c7} have been presented based on the Shannon entropy of random variables.
In recent years, Li et al.~\citep{structural-entropy, li2022principles} proposed an encoding-tree-based graph structural information metric, namely \textit{structural entropy}, to discover the natural hierarchical structure embedded in a graph.
The structural entropy has been used extensively in the fields of biological data mining~\citep{Hi-C, zhang2021supertad}, information security~\citep{REM, li2017resistance}, and graph neural networks~\citep{yang2023minimum, xu2022pooling, zou2023se}, etc.

\begin{figure*}[t]
\centering
\includegraphics[scale=0.54]{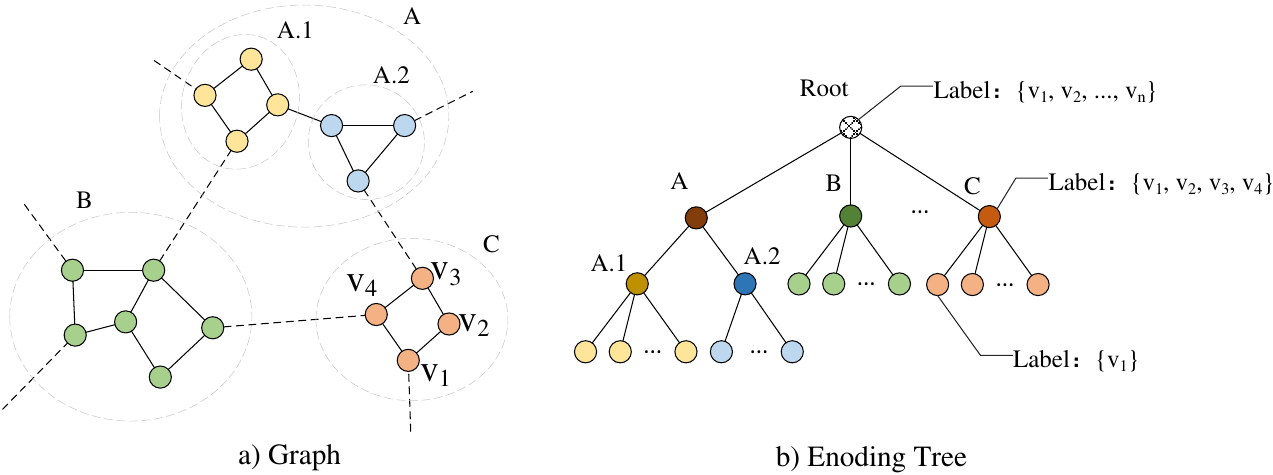}
\caption{
a) A graph containing three communities A, B, and C, where A is divided into two sub-communities A.1 and A.2. 
b) An encoding tree of the left graph. 
Each leaf node corresponds to a single graph node. 
Each branch node corresponds to a community. 
The root node corresponds to the graph node set.
}
\label{intro2in1}
\vspace{-0.5em}
\end{figure*}

The computation of structural entropy~\citep{structural-entropy} consists of three steps: encoding tree construction, node structural entropy calculation, and total structural entropy calculation.
Firstly, the graph node set is hierarchically divided into several communities (shown in Fig. \ref{intro2in1}(a)) to construct a partitioning tree, namely an encoding tree (shown in Fig. \ref{intro2in1}(b)). 
Secondly, the total node degree and cut edge number of each community are counted to compute the structural entropy of each non-root node in the encoding tree.
Thirdly, the structural entropy of the whole graph is calculated by summing up the node structural entropy.
In general, smaller structural entropy corresponds to better community partitioning.
Specifically, the minimized structural entropy, namely \textit{the graph structural entropy}, corresponds to the optimal encoding tree, which reveals the best hierarchical community partitioning of the graph.

In dynamic scenarios, a graph evolves from its initial state to many updated graphs during time series~\citep{harary1997dynamic}.
To efficiently measure the quality of evolving community partitioning, we need to incrementally compute the updated structural entropy at any given time.
Unfortunately, the current structural entropy methods~\citep{structural-entropy, li2022principles} do not support efficient incremental computation due to two challenges.
The first challenge is that the encoding tree needs to be reconstructed for every updated graph, which leads to enormous time consumption.
To address this issue, we propose two dynamic adjustment strategies for two-dimensional encoding trees\footnote{\revised{``Two-dimensional encoding tree'' means the height of the encoding tree is restricted to $2$. The corresponding structural entropy of the two-dimensional encoding tree is named ``two-dimensional structural entropy''.}}, namely \textit{the naive adjustment strategy} and \textit{the node-shifting adjustment strategy}.
The former strategy maintains the old community partitioning, and supports theoretical structural entropy analysis, while the latter dynamically adjusts the community partitioning by moving nodes between communities based on the principle of structural entropy minimization.
The second challenge is the high time complexity of structural entropy computation by the traditional definition.
To tackle this problem, we design an incremental framework, namely \textit{Incre-2dSE}, for efficiently measuring the updated two-dimensional structural entropy.
To be specific, \textit{Incre-2dSE} first utilizes the two dynamic adjustment strategies to generate \textit{Adjustments}, i.e., the changes of important statistics from the original graph to the updated graph and then uses the Adjustments to calculate the updated structural entropy by newly designed incremental formula.
\revised{Additionally, we also generalize our incremental methods to undirected weighted graphs and conduct a detailed discussion on the calculation of one-dimensional structural entropy for directed weighted graphs.}


We conduct extensive experiments on $3$ artificial dynamic graph datasets generated by \textit{Hawkes Process} and $3$ real-world datasets on the application of real-time monitoring of community partitioning quality (two-dimensional structural entropy) and community optimization.
Comprehensive experimental results show that our methods effectively capture the community evolution features and significantly reduce the time consumption with great interpretability.
All source code and data of this project are publicly available at \url{https://github.com/SELGroup/IncreSE}.

The main contributions of this paper are as follows:
\begin{itemize}
\item Proposing two dynamic adjustment strategies for two-dimensional encoding trees to avoid the reconstruction of the encoding tree for every updated graph.
\item Designing an incremental framework for efficiently measuring the updated two-dimensional structural entropy with low time complexity.
\item \revised{Extending the proposed methods to weighted graphs and providing new incremental computation methods for directed weighted graphs.}
\item Conducting extensive experiments on artificial and real-world datasets to demonstrate the effectiveness and efficiency of our method in dynamic measurement of community partitioning quality.
\end{itemize}

The article is structured as follows: 
Section~\ref{sec:problem} outlines the definitions and notations.
Section~\ref{sec:strategies} describes the dynamic adjustment strategies.
The algorithms are detailed in Section~\ref{sec:algo}, and \revised{Section~\ref{sec:ext} gives discussion on more complex graphs.}
The experiments are discussed in Section~\ref{sec:eval}.
Section~\ref{sec:relatedworks} presents the related works before concluding the paper in section~\ref{sec:conc}.

\section{Definitions and Notations}\label{sec:problem}
In this section, we summarize the notations in Table \ref{notations}, and formalize the definition of \textit{Incremental Sequence}, \textit{Dynamic Graph}, \textit{Encoding Tree}, and \textit{Structural Entropy} as follows. 

\begin{table}[]
\centering
\caption{Glossary of Notations.}
\begin{tabular}{@{}l|l@{}}
\toprule
\textbf{Notation} & \textbf{Description}                    \\ \midrule
$G$         & Graph       \\
$\mathcal{V}; \mathcal{E}$ & Node set; Edge set \\  
$v; e$            & Node; Edge                              \\
$d_i; d_v$               & Node degree of node $v_i$; Node degree of $v$               \\
$m$               & The total edge number of $G$\\ \midrule
$\mathcal{T}$               & Encoding tree                         \\
$\lambda$          & The root node of an encoding tree \\
$\alpha$          & The non-root node in an encoding tree, i.e., the community ID\\
$A$          & The set of all $1$-height nodes in an encoding tree\\
$T_\alpha$        & The label of $\alpha$, i.e., the node community corresponding to $\alpha$\\
$V_\alpha$        & The volume of $T_\alpha$ \\
$g_\alpha$        & The cut edge number of $T_\alpha$\\ \midrule
$\mathcal{G}$          & Dynamic graph \\
$G_0$       & Initial state of a dynamic graph\\
$G_t$ & The updated graph at time $t$\\
$\xi_t$ & Incremental sequence at time $t$\\
$\xi_{1 \rightarrow t}$ & Cumulative incremental sequence at time $t$\\
$n$ & Incremental size\\
$\delta(v)$ & The degree incremental of $v$ \\
$\delta_V(\alpha)$ & The volume incremental of $T_\alpha$ \\
$\delta_g(\alpha)$ & The cut edge number incremental of $T_\alpha$ \\ 
$\phi_\lambda$& The degree-changed node set \\ 
$\mathcal{A}$& The set of $1$-height tree nodes whose volume or cut edge \\
& number change \\ \midrule
$H^{\mathcal{T}}(G)$          & The structural entropy of $G$ by $\mathcal{T}$ \\
$H^\mathcal{T}_{GI}(G, n)$ & Global Invariant with incremental size $n$\\
$\Delta L$ & Local Difference, i.e., the approximation error between $H^\mathcal{T}(G)$ \\
& and $H^\mathcal{T}_{GI}(G, n)$\\ \bottomrule
\end{tabular}
\label{notations}
\end{table}

\begin{definition}[Incremental Sequence]
An incremental sequence is a set of incremental edges represented by $\xi = \{<(v_1,u_1), op_1>, <(v_2,u_2), op_2>, ..., <(v_n,u_n), op_n> \}$, where $(v_i,u_i)$ denotes an incremental edge $e_i$ with two endpoints $v_i$ and $u_i$.
The operator $op_i \in \{+, -\}$ represents that the edge $e_i$ will be added to or removed from a graph.
The number of the incremental edges $n$ is named the incremental size. 
\label{incre-seq}
\end{definition}

\begin{definition}[Dynamic Graph]
In this work, a dynamic graph is defined as a series of snapshots of a temporal, undirected, unweighted, and connected graph $\mathcal{G} = \{ G_0, G_1, ... , G_T\}$. 
$G_0=(\mathcal{V}_0,\mathcal{E}_0)$ denotes the initial state and $G_t=(\mathcal{V}_t,\mathcal{E}_t)$ denotes the updated graph at time $t$ $(1 \le t \le T)$.
To describe the temporal dynamic graph, we suppose that an incremental sequence $\xi_t$ arrives at each non-zero time $t$.
The updated graph $G_t$ is generated by orderly combining all new edges and nodes introduced by $\xi_t$ with $G_{t-1}$, i.e., $G_t := \text{CMB}(G_{t-1}, \xi_t)$. 
We further define the cumulative incremental sequence at time $t$, denoted by $\xi_{1\rightarrow t}$, as the sequence formed by sequentially concatenating sequences $\xi_1, \xi_2, ..., \xi_t$, and then we have $G_t := \text{CMB}(G_{0} , \xi_{1 \rightarrow t})$.
\label{dynam-graph}
\end{definition}

\begin{definition}[Encoding Tree]
The concept of the encoding tree is the same as ``the Partitioning Tree" proposed in the previous work~\citep{structural-entropy}.
The encoding tree $\mathcal{T}$ of a graph $G=(\mathcal{V}, \mathcal{E})$ is an undirected tree that satisfies the following properties:
\begin{enumerate}
\item The root node $\lambda$ in $\mathcal{T}$ has a label $T_\lambda=\mathcal{V}$.
\item Each non-root node $\alpha$ in $\mathcal{T}$ has a label $T_{\alpha} \subset \mathcal{V}$.
\item For each node $\alpha$ in $\mathcal{T}$, if $\beta_{1}, \beta_{2}, \ldots, \beta_{k}$ are all immediate successors of $\alpha$, then $T_{\beta_{1}}, \ldots, T_{\beta_{k}}$ is a partitioning of $T_{\alpha}$.
\item The label of each leaf node $\gamma$ is a single node set, i.e., $T_{\gamma}=\{v\}$.
\item $h(\alpha)$ denotes the height of a node $\alpha$ in $\mathcal{T}$. 
Let $h(\lambda)=0$ and $h(\alpha)=h\left(\alpha^{-}\right)+1$, where $\alpha^{-}$ is the parent of $\alpha$. 
The height of the encoding tree $h(\mathcal{T})$, namely the dimension, is equal to the maximum of $h(\gamma)$.
\end{enumerate}
\end{definition}

\begin{definition}[Structural Entropy]
The structural entropy is defined by Li and Pan~\citep{structural-entropy}. 
We follow this work and state the definition below.
Given an undirected, unweighted, and connected graph $ G = (\mathcal{V},\mathcal{E}) $ and its encoding tree $ \mathcal{T} $, the structural entropy of $ G $ by $ \mathcal{T} $ is defined as:
\begin{equation}
H^{\mathcal{T}}(G)=\sum_{\alpha \in \mathcal{T}, \alpha \neq \lambda} -\frac{g_{\alpha}}{2 m} \log \frac{V_{\alpha}}{V_{\alpha^-}},
\end{equation}
where $m$ is the total edge number of $G$; $g_{\alpha}$ is the cut edge number of $T_{\alpha}$, i.e., the number of edges between nodes in and not in $T_{\alpha}$; $V_{\alpha}$ is the volume of $T_{\alpha}$, i.e., the sum of the degrees of all nodes in $T_{\alpha}$; $\log(\cdot)$ denotes logarithm with a base of 2. 
We name $H^{\mathcal{T}}(G)$ as the $K$-dimensional structural entropy if $\mathcal{T}$'s height is $K$.

The graph structural entropy of $G$ is defined as:
\begin{equation}
H(G)=\min _{\mathcal{T}}\left\{H^{\mathcal{T}}(G)\right\},
\end{equation}
where $\mathcal{T}$ ranges over all possible encoding trees.

If the height of $\mathcal{T}$ is restricted to $K$, the $K$-dimensional graph structural entropy of $G$ is defined as:
\begin{equation}
H^K(G)=\min _{\mathcal{T}}\left\{H^{\mathcal{T}}(G)\right\},
\end{equation}
where $\mathcal{T}$ ranges over all the possible encoding trees of height $K$. 
The encoding tree corresponding to $H^K(G)$, which minimizes the $K$-dimensional structural entropy, is named the optimal $K$-dimensional encoding tree.
\end{definition}

\section{Dynamic Adjustment Strategies for Two-Dimensional Encoding Trees}\label{sec:strategies}
In this section, we first introduce the naive adjustment strategy and analyze the updated structural entropy under this strategy. 
Then, we describe the node-shifting adjustment strategy which leads to lower structural entropy, and provide its theoretical proof.
\revised{Finally, we give further discussion between the two proposed dynamic adjustment strategies.}

\subsection{Naive Adjustment Strategy}\label{sec:naive}
In this part, we first provide a formal description of the naive adjustment strategy.
Next, we introduce two metrics, Global Invariant and Local Difference,
to realize the incremental computation of updated structural entropy.
Finally, we analyze the boundedness and convergence of the Local Difference.

\subsubsection{Strategy Description}

\begin{figure}[h]
    \centering
    \includegraphics[width = 0.7\linewidth]{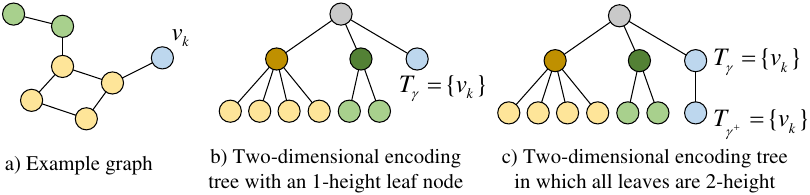}
    \caption{\revised{An example graph with its two equivalent two-dimensional encoding trees.}}
    \label{fig:eq-2dect}
\end{figure}

\newrevised{
Before we introduce the naive adjustment strategy, we first discuss the form of the two-dimensional encoding trees in detail and the formula of their corresponding structural entropy.}
\newrevised{
For all possible two-dimensional encoding trees, whenever there is a leaf node $\gamma$ ($T_\gamma = \{v_k\}$) whose height is $1$ (like Fig.~\ref{fig:eq-2dect}(b)), we can connect it with a child node $\gamma^+$ with the same label $T_{\gamma^+} = \{v_k\}$ to make all leaf nodes have height $2$ (Fig.~\ref{fig:eq-2dect}(c)).
At this time, the structural entropy remains unchanged since the additional term induced by $\gamma^+$, i.e. $-\frac{d_{k}}{2m}\log\frac{d_k}{d_k}$, equals to $0$.}

\newrevised{
In other words, the two encoding trees in Fig.~\ref{fig:eq-2dect} are equivalent.
Therefore, in this paper, we only consider the two-dimensional encoding trees where the height of all leaf nodes is $2$ for the sake of brevity.
Given a graph $G$ and its two-dimensional encoding tree $\mathcal{T}$, the two-dimensional structural entropy of $G$ by $\mathcal{T}$ can uniformly be formulated as:
}
\begin{equation}
\newrevised{
\begin{aligned}
H^\mathcal{T}(G)=\sum_{\alpha_i \in A}(-\frac{g_{\alpha_i}}{2m}\log\frac{V_{\alpha_i}}{2m}
+\sum_{v_j\in T_{\alpha_i}}-\frac{d_{j}}{2m}\log\frac{d_j}{V_{\alpha_i}}),
\end{aligned}
}
\label{2dSE}
\end{equation}
\newrevised{
where $A$ denotes the set of $1$-height nodes in $\mathcal{T}$, i.e., $A = \{\alpha \text{ in } \mathcal{T}| h(\alpha) = 1\}$.}

\begin{figure}[ht]
\centering
\includegraphics[scale=0.59]{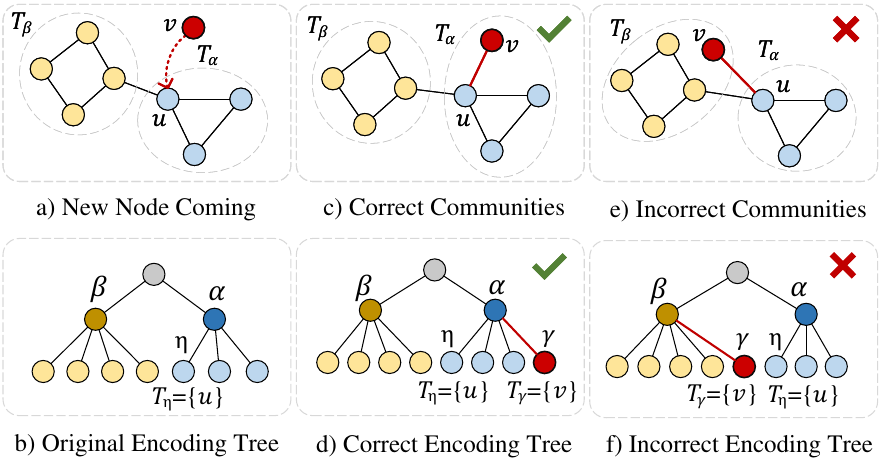}
\caption{
An example of the node strategy for adjusting two-dimensional encoding trees. 
}
\label{node-strategy}
\end{figure}

\newrevised{Now we present the description of the naive adjustment strategy.}
\newrevised{This} strategy comprises two parts: the edge strategy and the node strategy.
The edge strategy dictates that \textit{incremental edges do not alter the encoding tree's structure}.
On the other hand, the node strategy specifies that \textit{when a new node $v$ connects with an existing node $u$ (shown in Fig.~\ref{node-strategy}(a)), and $u$ corresponds to a leaf node $\eta$ in the two-dimensional encoding tree, i.e., $T_{\eta} = \{u\}$ (shown in Fig.~\ref{node-strategy}(b)), a new leaf node $\gamma$ with a label $T_\gamma = \{v\}$ will be set as an immediate successor of $\eta$'s parent ($\alpha$ in Fig.~\ref{node-strategy}(d)), instead of another 1-height node ($\beta$ in Fig.~\ref{node-strategy}(f))}.
We can describe the modification of the encoding trees from the community perspective.
Specifically, \textit{the incremental edges do not change the communities of the existing nodes} while \textit{the new node is assigned to its neighbor's community ($T_\alpha$ in Fig.~\ref{node-strategy}(c)), rather than another arbitrary community ($T_\beta$ in Fig.~\ref{node-strategy}(e))}.
Obviously, we can get the updated encoding tree, i.e., the updated community partitioning, in the time complexity of $O(n)$ given an incremental sequence of size $n$.
To ensure that the node strategy minimizes the updated structural entropy most of the time, we give the following theorem.
\begin{theorem}
Suppose that a new graph node $v$ is connected to an existing node $u$, where $\{ u\} \subseteq T_\alpha $.
If $\frac{2m+2}{V_\alpha+2} \ge e$, we have:
\begin{equation}
\begin{aligned}
H^{\mathcal{T}'}_{v\rightarrow \alpha}(G')  < H^{\mathcal{T}'}_{v\rightarrow \beta \neq \alpha}(G') ,
\end{aligned}
\label{node-entropy-difference}
\end{equation}
where $H^{\mathcal{T}'}_{v\rightarrow \alpha}(G')$ denotes the updated structural entropy when the new node $v$ is assigned to $u$'s community $T_\alpha$, i.e., $\{v\} \subset T_\alpha$, and $H^{\mathcal{T}'}_{v\rightarrow \beta \neq \alpha}(G')$ represents the updated structural entropy when $v$ is allocated to another arbitrary community $T_{\beta \neq \alpha}$, i.e., $ \{v\} \subset T_{\beta \neq \alpha}$.
\label{node strategy}
\end{theorem}
\textit{Proof.}
Differentiating the updated structural entropy of the two cases above, we can obtain:
\begin{align}
\Delta H^{\mathcal{T}'}(G') = &H^{\mathcal{T}'}_{v\rightarrow \alpha}(G') - H^{\mathcal{T}'}_{v\rightarrow \beta \neq \alpha}(G')\notag \\
=& \frac{1}{2m+2}[\log{\frac{V_\alpha+2}{2m+2}} + (g_{\alpha}-V_{\alpha})\log \frac{V_{\alpha}+1}{V_{\alpha}+2}-(g_{\beta}-V_{\beta})\log \frac{V_{\beta}}{V_{\beta}+1}].
\end{align}
Here we define:
\begin{align}
f_1(g, V) =& (g-V)\log \frac{V+1}{V+2}, \\
f_2(g, V)=& -(g-V)\log \frac{V}{V+1}.
\end{align}
Let $\theta $ $( 0 \le \theta <1)$ be the minimum proportion of the in-community edges in each community, i.e.,
\begin{align}
\theta = \min_\alpha \{ \frac{V_\alpha-g_\alpha}{V_\alpha}\}.
\end{align}
Since $1 \le V <2m$ and $0<g\le (1 - \theta) V$, we have:
\begin{align}
f_1(g, V) &< -V \log \frac{V+1}{V+2}<\frac{1}{\ln{2}}=\log{e},\\
f_2(g, V) &\le \theta V\log{\frac{V}{V+1}}\le \theta \log{\frac{1}{2}} = -\theta.
\end{align}
So
\begin{equation}
\begin{aligned}
\Delta H^{\mathcal{T}'}(G') < & \frac{1}{2m+2}(\log{\frac{V_\alpha+2}{2m+2}} + \log e - \theta)\\
= & \frac{1}{2m+2}\log( {\frac{V_\alpha+2}{2m+2}}\cdot 2^{-\theta} e).
\end{aligned}
\end{equation}
Therefore, if the following condition holds:
\begin{equation}
\begin{aligned}
\frac{2m+2}{V_\alpha+2} \ge \max\{2^{-\theta} e\}=e,\\
\end{aligned}
\end{equation}
then Eq. (\ref{node-entropy-difference}) holds, and thus Theorem~\ref{node strategy} is proven.\qed

According to Theorem~\ref{node strategy}, our node strategy minimizes the updated structural entropy when the total volume of the whole graph $2m$ is approximately larger than $e$ times the maximum volume $V_m$ of all communities.
Usually, we have $\theta \approx 1$, so the real condition is much looser.

\subsubsection{Global Invariant and Local Difference}
\newrevised{
In this part, we introduce two quantities, Global Invariant and Local Difference, to realize the approximation and the fast incremental calculation of the updated structural entropy by naive adjustment strategy.}
When an incremental sequence $\xi$ with size $n$ is applied to a graph $G$, resulting in a new graph $G'$ and its corresponding two-dimensional encoding tree $\mathcal{T}'$ using the naive adjustment strategy, the updated two-dimensional structural entropy can be expressed as:
\begin{equation}
\begin{aligned}
H^{\mathcal{T}'}(G')=&\sum_{\alpha_i \in A}(-\frac{g'_{\alpha_i}}{2m+2n}\log\frac{V'_{\alpha_i}}{2m+2n}+\sum_{v_j\in T_{\alpha_i}}-\frac{d'_{j}}{2m+2n}\log\frac{d'_j}{V'_{\alpha_i}}).
\end{aligned}
\label{2dSE-updated}
\end{equation}

\revised{
An intuitive way to calculate the updated two-dimensional structural entropy is to update the variables in Eq.~(\ref{2dSE}) and then compute via the updated formula Eq.~(\ref{2dSE-updated}).
However, the incremental size $n$ affects all terms in the summation equation in Eq.~(\ref{2dSE-updated}).
Therefore, the updating and calculation process costs at least $O(|\mathcal{V}|)$, which is huge when the graph becomes extremely large.
So how can we find an incremental formula with a smaller time complexity only related to the incremental size $n$?
An intuitive attempt is to make a difference between the updated structural entropy and the original one to try to compute the incremental entropy in $O(n)$.
Nevertheless, the fact that $m$ changes to $m+n$ in all terms of Eq.~(\ref{2dSE-updated}) makes it difficult to derive a concise formula of $O(n)$ from the difference equation.
}

\revised{To address this issue, we here introduce Global Invariant and Local Difference.
We define the Global Invariant (Definition~\ref{def:2dGI}) as an approximate updated structural entropy that only updates $m$ to $m+n$ in Eq.~(\ref{2dSE}) and keeps other variables unchanged.
The Local Difference (Definition~\ref{def:2dLD}) is defined as the difference between the updated structural entropy and the Global Invariant, which can also be regarded as the approximate error.}
\revised{Obviously,} we can get the Global Invariant in the time complexity of $O(1)$ if $S_N$, $S_C$, and $S_G$ are saved.
The Local Difference can also be computed in $O(n)$ given the necessary incremental changes.
Overall, the updated two-dimensional structural entropy can be calculated in $O(n)$ \revised{by computing and summing up the Global Invariant and the Local Difference.
In the experimental part, we use a more explicit and practiced form of the structural entropy update formula (Eq.~(\ref{eq:incre})) derived from the Global Invariant and the Local Difference.}

\begin{definition}[Global Invariant]
Given an original graph $G$ and its two-dimensional encoding tree $\mathcal{T}$, the Global Invariant is defined as an approximate value of the updated structural entropy after an incremental sequence with size $n$, i.e.,
\begin{align}
H_{GI}^\mathcal{T}(G,n) &= \sum_{\alpha_i \in A}(-\frac{g_{\alpha_i}}{2m+2n}\log\frac{V_{\alpha_i}}{2m+2n} + \sum_{v_j\in T_{\alpha_i}}-\frac{d_{j}}{2m+2n}\log\frac{d_j}{V_{\alpha_i}}) \notag\\ 
&= - \frac{1}{2m+2n}(S_N+S_C+S_G),
\label{2dGI}
\end{align}
where
\begin{align}
S_N &= \sum_{v_i \in \mathcal{V}}d_i\log{d_i},\\
S_C &= \sum_{\alpha_i \in A}(g_{\alpha_i}-V_{\alpha_i})\log V_{\alpha_i},\\
S_G &= -\sum_{\alpha_i \in A}g_{\alpha_i}\log(2m+2n).
\end{align}
\label{def:2dGI}
\end{definition}

\begin{definition}[Local Difference]
Given the updated graph $G'$, the updated two-dimensional encoding tree $\mathcal{T}'$, and incremental size $n$, the Local Difference is defined as the difference between the exact updated two-dimensional structural entropy and the Global Invariant, as shown below:
\begin{equation} 
\begin{aligned}
\Delta L = H^{\mathcal{T}'}(G')-H^\mathcal{T}_{GI}(G,n) = -\frac{1}{2m+2n}(\Delta S_N + \Delta S_C + \Delta S_G),
\end{aligned}
\label{2dLD}
\end{equation}
where
\begin{align}
\Delta S_N =& \sum_{v_k \in \phi_\lambda}[(d_k+\delta(v_k))\log(d_k+\delta(v_k)) - d_k\log{d_k}],\\
\Delta S_C =& \sum_{\alpha \in \mathcal{A}}[(g_{\alpha} +\delta_g(\alpha) - V_\alpha-\delta_V(\alpha))\log(V_{\alpha}+\delta_V(\alpha)) - (g_{\alpha}-V_\alpha)\log V_{\alpha}],\\
\Delta S_G =& -\sum_{\alpha \in \mathcal{A}}\delta_g(\alpha)\log(2m+2n).
\end{align}
Here, $\delta(v_k)$ denotes the incremental change in degree $d'_k - d_k$, $\delta_V(\alpha)$ represents the incremental change in volume $V'_\alpha-V_\alpha$, $\delta_g(\alpha)$ represents the incremental change in cut edge $g'_\alpha - g_\alpha$, $\phi_\lambda$ denotes the set of nodes that have changes in degree $\{v_k \in \mathcal{V}'| \delta(v_k) \neq 0\}$, and $\mathcal{A}$ denotes the set of $1$-height tree nodes that have changes in $V_\alpha$ or $g_\alpha$, i.e., $\mathcal{A} = \{\alpha \in A' | \delta_V(\alpha) \neq 0 \text{ or } \delta_g(\alpha) \neq 0\}$.
\label{def:2dLD}
\end{definition}

\subsubsection{Boundedness Analysis}
According to Eq.~(\ref{2dLD}), the bounds of $\Delta L$ can be obtained by analyzing its components, namely $\Delta S_N$, $\Delta S_C$ and $\Delta S_G$.
First, we analyze the maximum and minimum values of $\Delta S_N$.
We define
\begin{equation}
s_N(d, x) = (d+x)\log(d+x) - d\log{d}.
\end{equation}
Since $s_N(d,n)$ is monotonically increasing with $d$, $\Delta S_N$ takes the maximum value when $n$ new incremental edges connect the two nodes with the largest degree.
Therefore, we have
\begin{equation}
\Delta S_N \le 2s_N(d_{m},n),
\end{equation}
where $d_m$ denotes the maximum degree in $G$.
Since multiple edges are not allowed, the equality may hold if and only if $n = 1$.
When each of the $n$ incremental edges connects a one-degree node and a new node, $\Delta S_N$ is minimized:
\begin{equation}
\Delta S_N \ge ns_N(1,0).
\end{equation}
Second, we analyze the bounds of $\Delta S_C$ and $\Delta S_G$.
For convenience, we define
\begin{equation}
\begin{aligned}
 \Delta S_{CG} = \Delta S_C +\Delta S_G.
\end{aligned}
\end{equation}
We commence by analyzing the bound of $\Delta S_{CG}$ when adding one new edge.
If a new edge is added between two communities $T_{\alpha_1}$ and $T_{\alpha_2}$, we can get
\begin{align}
\Delta S_{CG} =& (g_{\alpha_1} - V_{\alpha_1})\log(V_{\alpha_1} + 1) - (g_{\alpha_1}-V_{\alpha_1})\log V_{\alpha_1} \notag \\
&+ (g_{\alpha_2} - V_{\alpha_2})\log(V_{\alpha_2} + 1) - (g_{\alpha_2}-V_{\alpha_2})\log V_{\alpha_2} - 2 \log (2m+2).
\end{align}
Thus we have
\begin{align}
\Delta S_{CG} \ge &  2V_m\log(\frac{V_m}{V_m + 1}) - 2 \log (2m+2) , \end{align}
and
\begin{align}
\Delta S_{CG} \le & -2 \log(2m+2),
\end{align}
where $V_m$ denotes the maximum volume of all $T_\alpha$.
If a new edge is added within a single community $T_\alpha$ (or a new node is connected with an existing node in $T_\alpha$), we have
\begin{equation}
\begin{aligned}
\Delta S_{CG} =& (g_\alpha  - V_\alpha -2)\log(V_\alpha+2) - (g_\alpha-V_\alpha)\log V_\alpha.\\
\end{aligned}
\end{equation}
Then we can obtain
\begin{align}
\Delta S_{CG} \ge & - (V_m + 2)\log(V_m+2) + V_m\log V_m,
\end{align}
and
\begin{align}
\Delta S_{CG} \le& -2\log(V_{min}+2),
\end{align}
where $V_{min}$ denotes the minimum volume of all $T_\alpha$.
We next analyze the bound of $\Delta S_{CG}$ when adding $n$ new edges.
When the $n$ edges are all between the two communities with the largest volume, we have:
\begin{equation}
\begin{aligned}
\Delta S_{CG} \ge &  2V_m\log(\frac{V_m}{V_m + n}) - 2n \log (2m+2n)\\
 >  &- 2n- 2n \log (2m+2n),\\
\end{aligned}
\end{equation}
and $\Delta S_{CG}$ takes the minimum value:
\begin{equation}
\begin{aligned}
\Delta S_{CGmin} =  &- 2n- 2n \log (2m+2n).\\
\end{aligned}
\end{equation}
When each of the $n$ edges is added within $n$ communities with the smallest volume, respectively, $\Delta S_{CG}$ takes its maximum value:
\begin{equation}
\begin{aligned}
\Delta S_{CGm} = & -2n\log(V_{min}+2).
\end{aligned}
\end{equation}
Finally, we can get a lower bound of $\Delta L$ as
\begin{align}
\text{LB}(\Delta L) =& -\frac{1}{2m+2n}(2s_N(d_m,n) + \Delta S_{CGm}) \notag \\
=&\frac{1}{m+n}[d_m\log d_m-(d_m+n)\log{(d_m+n)} +n \log(V_{min}+2)].
\label{lb2}
\end{align}
An upper bound of $\Delta L$ is as follows:
\begin{align}
\text{UB}(\Delta L) & =  -\frac{1}{2m+2n}(ns_N(1,0) + \Delta S_{CGmin}) \notag \\
&= \frac{ n\log(m+n) + \frac{5}{2}n}{m+n}.
\label{ub2}
\end{align}
\revised{\textit{Discussion:
The boundedness analysis gives the lower and upper bound of the Local Difference.
This suggests that when we compute Global Invariant to quickly get an approximate value of the updated structural entropy, the approximate error is bounded and decreases as the graph grows larger and thus the validity and accuracy of the approximation are guaranteed.
}}

\subsubsection{Convergence Analysis}
In this section, we analyze the convergence of the Local Difference as well as its first-order absolute moment.
To denote that one function converges at the same rate or faster than another function, we use the notation $g(m) = O(f(m))$, which is equivalent to $\lim_{m\rightarrow \infty}\frac{g(m)}{f(m)} = C$, where $C$ is a constant.
\begin{theorem}
Given the incremental size $n$, the Local Difference converges at the rate of $O(\frac{\log m}{m})$, represented as:
\begin{equation}
\begin{aligned}
\Delta L =& O(\frac{\log m}{m}).
\end{aligned}
\end{equation}
\label{conv-L}
\end{theorem}
\textit{Proof.}
The lower bound of $\Delta L$ is given by:
\begin{align}
\text{LB}(\Delta L) =&\frac{d_m\log d_m-(d_m+n)\log{(d_m+n)}}{m+n}+\frac{n \log(V_{min}+2)}{m+n} \notag\\
\ge&\frac{m\log m-(m+n)\log{(m+n)}}{m+n} +\frac{n \log(2+2)}{m+n}, \notag\\
\ge & \frac{1}{m+n}[\log (1-\frac{n}{m+n})^m - n\log (m+n)] \notag\\
= & O(\frac{\log m}{m}).
\end{align}
Similarly, the upper bound is given by:
\begin{align}
\text{UB}(\Delta L) =& \frac{ n\log(m+n) + \frac{5}{2}n}{m+n}  \notag\\
=& O(\frac{\log m}{m}).
\end{align}
Since
\begin{align}
 \text{LB}(\Delta L) \le \Delta L \le \text{UB}(\Delta L),
\end{align}
Theorem \ref{conv-L} is proved.\qed

It follows that the difference between the exact value of the updated two-dimensional structural entropy and the Global Invariant converges at the rate of $O(\frac{\log m}{m})$. 

\begin{definition}
Let $X$ be a random variable representing the incremental size $n$. We remind that $\mathbb{E}[X] = \overline n$.
\end{definition}
\begin{theorem}
The first-order absolute moment of the Local Difference converges at the rate of $O(\frac{\log m}{m})$:
\begin{equation}
\begin{aligned}
\mathbb{E}[|\Delta L| ] = O(\frac{\log m}{m}).
\end{aligned}
\end{equation}
\label{first-order}
\end{theorem}
\textit{Proof.}
We can represent the expectation of the lower bound of $\Delta L$ as:
\begin{align}
\mathbb{E}[|\text{LB}(\Delta L)|] = &  \mathbb{E}[|\frac{d_m\log d_m-(d_m+X)\log(d_m+X)}{m+X} +\frac{n\log(V_{min}+2)}{m+X}|] \notag\\
\le & \mathbb{E}[\frac{(m+X)\log{(m+X)}-m\log m}{m+X}] +\mathbb{E}[\frac{X\log(m+2)}{m+X}] \notag\\
\le & \frac{m\log m-(m+\overline n)\log{(m+\overline n)}}{m+\overline n} + \frac{\overline n\log(m+2)}{m+\overline n} \notag\\
= & O(\frac{\log m}{m}).
\end{align}
Similarly, the expectation of the upper bound is given by:
\begin{align}
\mathbb{E}[|\text{UB}(\Delta L)|] = &  \mathbb{E}[\frac{ X\log(m+X) + \frac{5}{2}X}{m+X}] \notag\\
\le & \frac{ \overline n\log(m+ \overline n) + \frac{5}{2}\overline n}{m+\overline n} \notag\\
= & O(\frac{\log m}{m}).
\end{align}
Finally, since
\begin{equation}
\begin{aligned}
0 \le \mathbb{E}[|\Delta L|] \le \max\{ \mathbb{E}[|\text{LB}(\Delta L)|], \mathbb{E}[|\text{UB}(\Delta L)|]\},
\end{aligned}
\end{equation}
Theorem \ref{first-order} is proved.\qed

\revised{\textit{
Discussion:
The convergence analysis gives proof of the convergence of the Local Difference and its first-order absolute moment.
That is, when we use Global Invariant to approximate the updated structural entropy, the approximate error and its absolute value's expectation both converge to $0$ no slower than $O(\frac{\log m}{m})$ as the graph edge number $m$ grows larger, suggesting a high level of confidence that our approximation is reliable.
Additionally, the convergence analysis also demonstrates that the updated structural entropy is mainly contributed by the incremental size $n$ other than the position of the incremental edges when the graph is extremely large.
It is because when the graph grows larger, we can approximate the updated structural entropy with an approximate error that converges to $0$ by simply changing $m$ to $m+n$.}
}
\revised{
\subsubsection{Limitations}\label{sec:na-lim}
The limitations of the naive adjustment strategy are listed below.
First, this strategy cannot deal with multiple incremental edges at the same time, e.g. a new node appears connecting with two different existing nodes.
An alternative solution is to arrange all incremental edges at a certain time stamp into a sequence with which we can add the edges one by one while keeping the connectivity of the graph. 
In this way, the community of newly introduced nodes is inevitably related to the input order of the incremental edges.
Second, it cannot handle edge or node deletions.
Third, the community of the existing nodes remains unchanged, which is sub-optimal in most cases. 
}

\subsection{Node-Shifting Adjustment Strategy}
Although the naive adjustment strategy can quickly obtain an updated two-dimensional encoding tree and its corresponding structural entropy, we still need a more effective strategy to get a better community partitioning towards lower structural entropy.
Therefore, we propose another novel dynamic adjustment strategy, namely \textit{node-shifting}, by moving nodes to their \textit{optimal preference communities} (Definition~\ref{def:OPC}) iteratively.
Different from the naive adjustment strategy, edge changes can change the communities of the existing nodes to minimize the structural entropy.
Besides, this strategy supports multiple incremental edges at the same time and the removal of the existing edges.
\newrevised{Therefore, the node-shifting adjustment strategy fully overcomes the limitations of the naive adjustment strategy listed in Section~\ref{sec:na-lim}.}
In the following, we first describe the node-shifting adjustment strategy in detail and then prove that the node's movement towards its optimal preference community can get the lowest structural entropy greedily.
Finally, we discuss the limitations of this strategy.

\subsubsection{Strategy Description}
We first define the optimal preference community (OPC) (Definition~\ref{def:OPC}) \revised{as the best community for a target node, i.e., if the target node moves into its OPC, the overall two-dimensional structural entropy must be the lowest compared to other community other than OPC.}
Then the node-shifting adjustment strategy can be described as follows:
\textit{(1) let involved nodes be all nodes that appeared in the incremental sequence;
(2) for each involved node, move it to its OPC;
(3) update the involved nodes to all nodes connected with the shifted nodes but in different communities, then repeat step (2).}

\begin{definition}[Optimal Preference Community (OPC)]
Given a graph $G = (\mathcal{V}, \mathcal{E})$ and a target node $v_t \in \mathcal{V}$, the optimal preference community of $v_t$ is defined as the community $T_{\alpha^*}$ in which
\begin{equation}
\begin{split}
\alpha^*= \left \{
\begin{array}{ll}
\arg \min_\alpha [(g_\alpha - V_\alpha)\log\frac{V_\alpha}{V_\alpha+d_t} + 2d^{(\alpha)}\log\frac{V_\alpha + d_t}{2m}], & v_t \notin T_\alpha;\\
\arg \min_\alpha [(g_\alpha - V_\alpha + d_t + d^{(\alpha)})\log\frac{V_\alpha-d_t}{V_\alpha} + 2d^{(\alpha)}\log\frac{V_\alpha}{2m}], & v_t \in T_\alpha,\\
\end{array}
\right.
\end{split}
\label{eq:OPC}
\end{equation}
where $d^{(\alpha)}$ denotes the number of the edges connected between $v_t$ and $T_{\alpha}$.
\label{def:OPC}
\end{definition}

\begin{figure}[t]
    \centering
    \includegraphics[width = \linewidth]{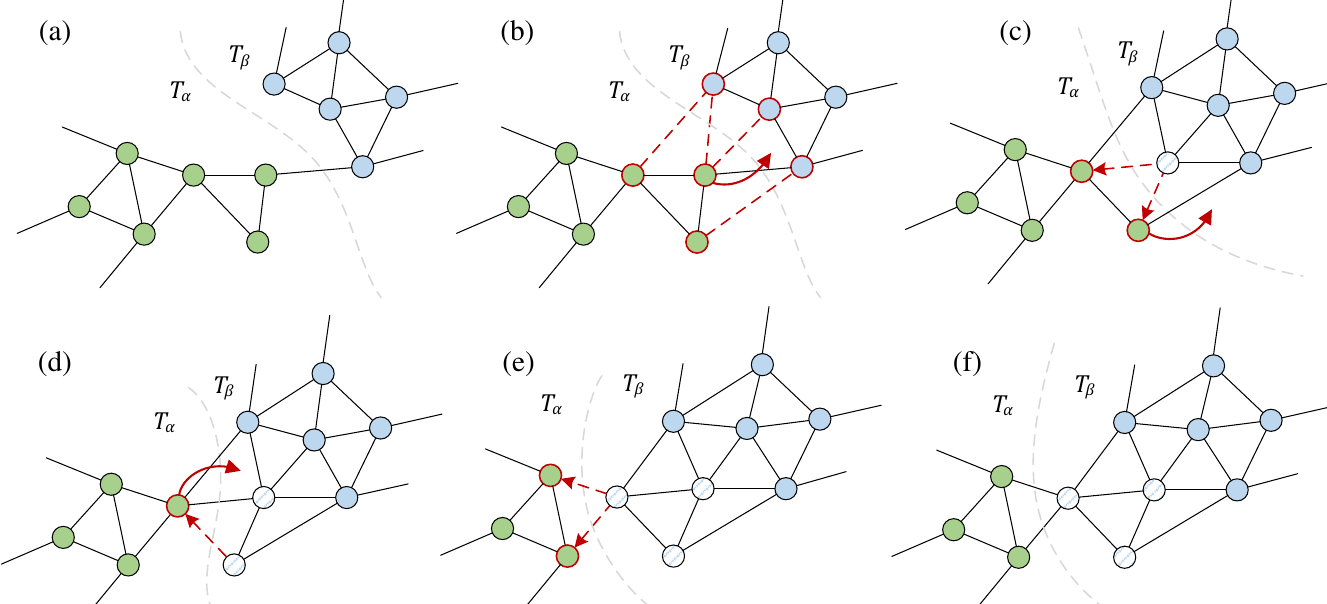}
    \caption{An example of the node-shifting adjustment strategy for adding new edges.}
    \label{fig:ns-ex}
\end{figure}

\begin{figure}[t]
    \centering
    \includegraphics[scale=0.60]{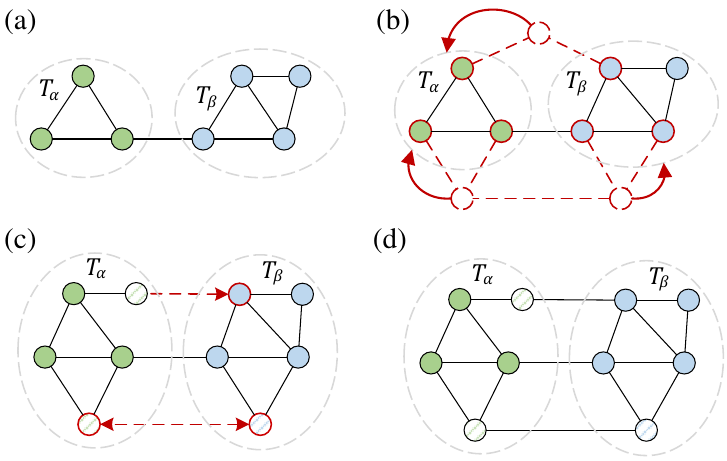}
    \caption{An example of the node-shifting adjustment strategy for adding new nodes.}
    \label{fig:ns-ex2}
\end{figure}

Fig.~\ref{fig:ns-ex} and Fig.~\ref{fig:ns-ex2} give examples to illustrate the node-shifting adjustment strategy in different situations.
Fig.~\ref{fig:ns-ex} shows how incremental edges affect community partitioning.
In Fig.~\ref{fig:ns-ex}(a), the graph is divided into $2$ communities $T_\alpha$ and $T_\beta$.
In Fig.~\ref{fig:ns-ex}(b), $4$ incremental edges (red dotted) are inserted into the graph.
Then all involved nodes (red outlined) are checked for moving into their OPCs.
In this step, one green node is shifted from $T_\alpha$ to $T_\beta$ (denoted by the red arrow).
In Fig.~\ref{fig:ns-ex}(c), the shifted node in the previous step ``sends messages'' (red dotted arrows) to its neighbors in $T_\alpha$  (red outlined).
The nodes that received the message (red outlined) are then checked for shifting.
At this time, another green node moves into $T_\beta$.
In Fig.~\ref{fig:ns-ex}(d)-(e), the graph follows the above process to continue the iterative update.
The final community partitioning is shown in Fig.~\ref{fig:ns-ex}(f).
Fig.~\ref{fig:ns-ex2} shows how new nodes are assigned to communities.
Fig.~\ref{fig:ns-ex2}(a) gives a $7$ nodes graph with $2$ communities.
In Fig.~\ref{fig:ns-ex2}(b), $3$ new nodes (white filled) are added with $7$ incremental edges and they belong to no community.
Then all of them and their existing neighbors become involved nodes.
Next, the upper new node is assigned to $T_\alpha$ because $T_\alpha$ is determined as its OPC.
Also, the lower two new nodes are moved into their OPCs.
In Fig.~\ref{fig:ns-ex2}(c), the new involved nodes (red outlined) are checked.
Fig.~\ref{fig:ns-ex2}(d) shows the final state of this node-shifting process.

\subsubsection{Theoretical Proof}
\begin{figure}
    \centering
    \includegraphics[scale = 0.75]{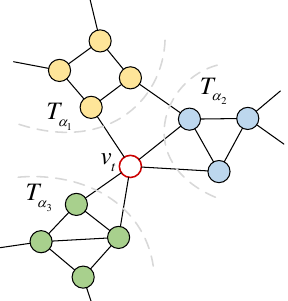}
    \caption{A simplified model for theoretical analysis.}
    \label{fig:simp}
\end{figure}

In this part, we provide a simplified model \newrevised{(Fig.~\ref{fig:simp})} to theoretically derive the OPC's solution formula (Eq.~(\ref{eq:OPC})).
In the graph of this model, there exists $r$ communities $T_{\alpha_1}, T_{\alpha_2}, ..., T_{\alpha_r}$.
There is also a target node $v_t$ which does not belong to any community.
The number of the edges connected between $v_t$ and $T_{\alpha_i}$ is denoted by $d^{(i)}$.
The volume and the number of the cut edges of $T_{\alpha_i}$ are denoted by $V_i$ and $g_i$, respectively.
Then we have Theorem~\ref{theo:OPC}.

\begin{theorem}

Suppose that the node $v_t$ is moving into community $T_{\alpha_k}, k \in \{1,2,...,r\}$.
The updated structural entropy is minimized when $v_t$ moves into $T_{\alpha_{k^*}}$ where 
\begin{equation}
\begin{aligned}
k^* = \arg \min_k [(g_k - V_k)\log\frac{V_k}{V_k+d_t} + 2d^{(k)}\log\frac{V_k + d_t}{2m}].
\label{eq:OPC-brief}
\end{aligned}
\end{equation}
\label{theo:OPC}
\end{theorem}

\textit{Proof.}
Let $H_k$ be the two-dimensional structural entropy after $v_t$ moves into $T_{\alpha_k}$.
Then $H_k$ is given by
\begin{equation}
\begin{aligned}
H_k = & \sum_{\alpha_i \neq \alpha_k}(-\frac{g_i}{2m}\log\frac{V_i}{2m} + \sum_{v_j\in T_{\alpha_i}} - \frac{d_j}{2m}\log \frac{d_j}{V_i}) + (-\frac{g_k+d_t-2d^{(k)}}{2m} \log \frac{V_k+d_t}{2m} \\
&+\sum_{v_q \in T_{\alpha_k}/\{v_t\}} - \frac{d_q}{2m}\log\frac{d_q}{V_k+d_t}-\frac{d_t}{2m}\log\frac{d_t}{V_k+d_t}).\\
\end{aligned}
\end{equation}
Therefore, the structural entropy is minimized when $v_t$ moves into $T_{\alpha_{k^*}}$ where
\begin{equation}
\begin{aligned}
k^* = \arg \min_k H_k.\\
\end{aligned}
\end{equation}
Let the structural entropy before the node movement be $\tilde H$ given by
\begin{equation}
\begin{aligned}
\tilde H = \sum_{\alpha_i}(-\frac{g_i}{2m}\log\frac{V_i}{2m} + \sum_{v_j\in T_{\alpha_i}} - \frac{d_j}{2m}\log \frac{d_j}{V_i}) + (-\frac{d_t}{2m}\log\frac{d_t}{2m}-\frac{d_t}{2m}\log\frac{d_t}{d_t}).\\
\end{aligned}
\end{equation}
Since $\tilde H$ is independent of $k$, we have
\begin{equation}
\begin{aligned}
k^* =&\arg \min_k 2m(H_k - \tilde H) \\
=&\arg \min_k [(g_k - V_k)\log\frac{V_k}{V_k+d_t} + 2d^{(k)}\log\frac{V_k + d_t}{2m}].\\
\end{aligned}
\end{equation}
Therefore Theorem~\ref{theo:OPC} is proved.\qed

In practice, all nodes belong to their communities.
We can first move the target node out of its community, and then use Eq.~(\ref{eq:OPC-brief}) to determine the OPC.
This process is equivalent to directly using Definition~\ref{def:OPC} without moving out the target node.

\subsubsection{Limitations}
\begin{figure}
    \centering
    \includegraphics[scale = 0.7]{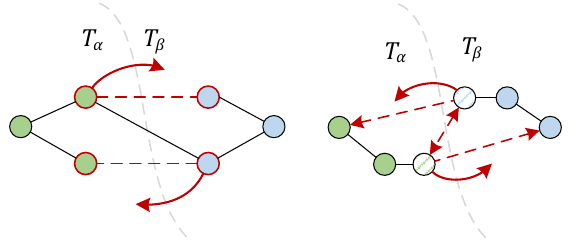}
    \caption{
    An example where the node-shifting adjustment strategy does not converge.
    The left is a graph added with two incremental edges which cause two nodes to shift.
    The right shows the updated communities after the first iteration and the future movement at the second iteration.
    After the second iteration, the graph becomes the left again.
    }
    \label{fig:not_converge}
\end{figure}

The limitations of the node-shifting adjustment strategy are listed below.
\revised{First}, it is hard to give the bound of the gap between the Global Invariant and the updated structural entropy \revised{for further theoretical analysis}.
\revised{Second}, the node-shifting adjustment strategy may not converge in some cases (Fig.~\ref{fig:not_converge} gives an example), which forces us to set the maximum number of iterations.

\subsection{\revised{Further Discussion between the Two Dynamic Adjustment Strategies}}
\revised{
\textbf{Similarities:}
(1) Both dynamic adjustment strategies are designed to incrementally change the original two-dimensional encoding trees to adapt the incremental edges and nodes in dynamic scenarios.
(2) The time complexities for computing the updated structural entropy of both strategies are significantly lower than the original calculation formula (detailed analysis is shown in Section~\ref{sec:incre}).
(3) Both strategies cannot handle the birth of new communities and the dismission of the existing communities.
}

\revised{
\textbf{Differences:}
(1) The focuses of the two strategies are different.
The naive adjustment strategy emphasizes theoretical analysis, such as boundedness and convergence analysis, and acts as a fast incremental baseline in experimental evaluations.
By contrast, the node-shifting adjustment strategy mainly focuses on addressing the limitations of the naive strategy (Section~\ref{sec:na-lim}) and the dynamic optimization of the existing communities towards a lower structural entropy.
(2)	The ways of updating encoding trees, or updating community partitionings, of these two strategies are different. 
In the naive adjustment strategy, new edges do not change the communities of the existing nodes, and new nodes are assigned to the direct neighbors’ communities. 
While in the node-shifting adjustment strategy, the influence on community adjustment of new edges is considered and the new nodes’ community is also determined by the incremental edges.
(3) The time complexity of the naive adjustment strategy is fixed while that of the node-shifting strategy grows nearly linearly with the iteration number $N$.
Experiments show that the naive strategy is faster than the node-shifting strategy with $N\ge5$ in most cases (Fig.~\ref{fig:exp3}).
}

\section{Incre-2dSE: an Incremental Measurement Framework of the Updated Two-Dimensional Structural Entropy}\label{sec:algo}

\subsection{
\revised{Definitions}
}
\revised{
In this part, we present the definitions of Structural Data, Structural Expressions, and Adjustment, which will be employed in subsequent sections.
}

\begin{definition}[Structural Data]
Given a graph $G$, the Structural Data of $G$ is defined as follows:
\begin{enumerate}
\item (node level) the degree $d_i$ of each $v_i \in \mathcal{V}$; 
\item (community level) the volume $V_\alpha$ and the cut edge number $g_\alpha$;
\item (graph level) the total edge number $m$;
\item (node-community map) the community ID $v_i$ belongs to, denoted by $\alpha(v_i) \in A$ where $v_i \in T_{\alpha(v_i)}$;
\item (community-node map) the community $T_\alpha$ of each $\alpha \in A$.
\end{enumerate}
\label{def:sd}
\end{definition}

\begin{definition}[Structural Expressions]
The Structural Expressions of $G$ are defined as follows:
\begin{enumerate}
\item (node level) 
\begin{equation}
\hat{S}_{N}=\sum_{d \in D}k_d d\log d,
\label{2dSN}
\end{equation}
where $k_d$ denotes the node number of each $d \in D$ while $D$ denotes the set of all distinct degrees in $G$;
\item (community level)
\begin{equation}
\hat{S}_{C}= \sum_{\alpha \in A}(g_{\alpha}-V_{\alpha})\log V_{\alpha};
\label{2dSC}
\end{equation}
\item (graph level) 
\begin{equation}
\hat{S}_{G} = -\sum_{\alpha \in A} g_{\alpha}.
\label{2dSG}
\end{equation}
\end{enumerate}
\label{def:se}
\end{definition}

\begin{definition}[Adjustment]
The Adjustment from the original graph $G$ to the updated graph $G'$ is defined as follows:
\begin{enumerate}
\item (node level) the degree change $\delta(v)$ for each node $v \in \phi_\lambda$ and the node number change $\delta_k (d) = k'_d - k_d$ of each $d \in \mathcal{D}$, where $\mathcal{D}$ denotes the set of the degrees which have node number changes from $G$ to $G'$;
\item (community level) the volume change $\delta_V(\alpha)$ and the cut edge number change $\delta_g(\alpha)$ of each $\alpha \in \mathcal{A}$;
\item (graph level) the total edge number change $n$;
\item (node-community map) the change list of the node-community map Structural Data denoted by $J_{n-c} = \{...,(v_i, \alpha'(v_i)),...\}$ where $\alpha'(v_i)$ denotes the new community ID of $v_i$;
\item (community-node map) the change list of the community-node map Structural Data denoted by $J_{c-n} = \{...,(\alpha_i, v_j, +/-),...\}$ where $(\alpha_i, v_j,$ $ +/-)$ denotes that community $T_{\alpha_i}$ is updated as $T_{\alpha_i} \cup \{v_j\}$ or $T_{\alpha_i} / \{v_j\}$.
\end{enumerate}
\label{def:adj}
\end{definition}

\subsection{Outline}

\begin{figure*}[ht]
\centering
\includegraphics[width = \linewidth]{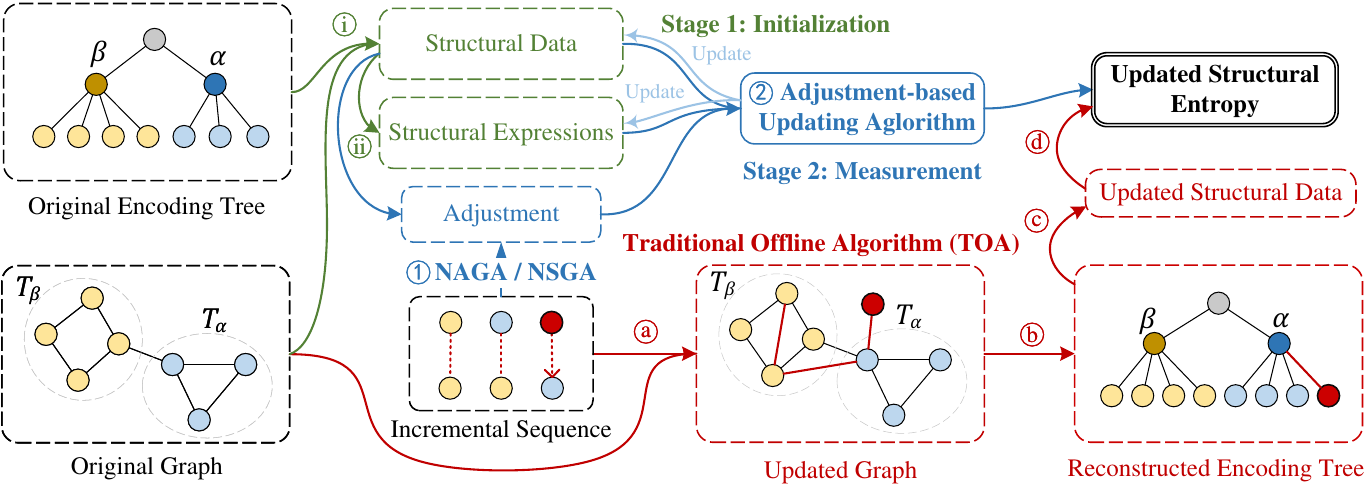}
\caption{
The outline of \textit{Incre-2dSE} (including two stages, Initialization and Measurement) and the traditional offline algorithm.}
\label{outline}
\end{figure*}

The illustration of our incremental framework \textit{Incre-2dSE} and its static baseline, the traditional offline algorithm (TOA), is shown in Fig.~\ref{outline}.
\textit{Incre-2dSE} aims to \revised{efficiently} measure the updated two-dimensional structural entropy \revised{while dynamically adjusting the community partitioning} given the original graph, the original encoding tree, and the incremental sequences.
This framework consists of two stages, \textit{initialization} and \textit{measurement}.
In the initialization stage, the Structural Data (Definition~\ref{def:sd}), which contains a graph's essential data to compute the structural entropy, is extracted from the original graph and its encoding tree (Fig.~\ref{outline}\circled{i}).
Then the Structural Expressions (Definition~\ref{def:se}), which are defined as the expressions of the Structural Data, are computed and saved (Fig.~\ref{outline}\circled{ii}).
For the same original graph, Initialization only needs to be performed once.
In the measurement stage, the Adjustment (Definition~\ref{def:adj}), which is defined as a data structure storing the changes in degree, volume, and cut edge number from the original graph to the updated graph, is first generated and saved according to the structural data and the incremental sequence by the Adjustment generation algorithm with the naive adjustment strategy (NAGA) or the node-shifting adjustment strategy (NSGA) (Fig.~\ref{outline}\circled{1}).
Then, the Adjustment-based incremental updating algorithm (AUIA) is called to gather the Structural Data, the Structural Expression, and the Adjustment to efficiently calculate the updated structural entropy and update the Structural Data and the Structural Expressions (Fig.~\ref{outline}\circled{2}).
As the baseline, TOA commences by updating the graph using the incremental sequence (Fig.~\ref{outline}\textcircled{a}).
Next, the new encoding tree of the updated graph is reconstructed using a static community detection method (Fig.~\ref{outline}\textcircled{b}).
Then, the updated Structural Data is extracted (Fig.~\ref{outline}\textcircled{c}), and finally, the updated structural entropy is computed by definition (Fig.~\ref{outline}\textcircled{d}).

\subsection{The Incremental Framework}\label{sec:incre}
\subsubsection{Stage 1: Initialization}

Given a graph $G = (\mathcal{V}, \mathcal{E})$ as a sparse matrix and its two-dimensional encoding tree represented by a dictionary like \{community ID $1$: node list $1$, community ID $2$: node list $2$, ...\}, the Structural Data (Definition~\ref{def:sd}) can be easily obtained and saved in the time complexity of $O(|\mathcal{E}|)$ (Fig.~\ref{outline}\circled{i}).
Then the Structural Expressions (Definition~\ref{def:se}) can be calculated with the saved Structural Data in $O(|\mathcal{V}|)$ (Fig.~\ref{outline}\circled{ii}).
Overall, the Initialization stage requires total time complexity $O(|\mathcal{E}|)$.

\subsubsection{Stage 2: Measurement}
In this stage, we first need to generate the Adjustment (Definition~\ref{def:adj}) from $G$ to $G'$.
We provide two algorithms for generating Adjustments by the proposed two dynamic adjustment strategies, namely \textit{the naive adjustment generation algorithm (NAGA)} and \textit{the node-shifting adjustment generation algorithm (NSGA)} (Fig.~\ref{outline}\circled{1}).
The input of both two algorithms is the Structural Data of the original graph and an incremental sequence and the output is an Adjustment.
The pseudo-code of \textit{NAGA} and \textit{NSGA} are shown in Algorithm~\ref{algo:NAGA} and Algorithm~\ref{algo:NSGA}, respectively.
The time complexity of \textit{NAGA} is $O(n)$ because it needs to traverse $n$ edges in the incremental sequence and it only costs $O(1)$ for each edge.
In \textit{NSGA}, we first need $O(n)$ to initialize the Adjustment (line $5$-$31$).
Second, in the node-shifting part (line $32$-$51$), we need to determine the OPC for all $|I|$ involved nodes, which costs $O(|A||I|)$.
This step is repeated $N$ times and the time cost is $O(|A|(|I_1|+|I_2|+...+|I_N|))$, where $I_i$ denotes the number of the involved nodes in the $i$-th iteration.
Since $|I_1| \le n$ and $|I_{i+1}|\le |I_i|$ most of the time, the total time complexity of \textit{NSGA} is $O(nN|A|)$.

\begin{algorithm}[]
    \fontsize{11}{12}\selectfont
  \SetAlgoLined
  \SetKwInOut{Input}{Input}\SetKwInOut{Output}{Output}
  \SetKwFunction{GetLength}{GetLength}
  \SetKwFunction{NULL}{NULL}
  \SetKwFunction{Partition}{Partition}
  \Input{The Structural Data ($d_i$, $V_\alpha$, $g_\alpha$, $m$, $\alpha(v_i)$, and $T_\alpha$) of $G$, and an incremental sequence $\xi$ from $G$ to $G'$.}
  \Output{The Adjustment ($\delta(v_i)$, $\delta_k(d)$, $\delta_V(\alpha)$,  $\delta_g(\alpha)$, $n$, $J_{n-c}$, and $J_{c-n}$) from $G$ to $G'$ by the naive adjustment strategy.}
    $n := \GetLength(\xi)$;\\
    $\delta(v_i):=0$, $\delta_k(d):=0$, $\delta_V(\alpha):=0$,  $\delta_g(\alpha):=0$, $\mathcal{D}:=\emptyset$, $\mathcal{A} := \emptyset$, $J_{n-c}:=\emptyset$, $J_{c-n}:=\emptyset$\;
    Let the proxy maps be $\hat \alpha(v_i) := \alpha(v_i), v_i \in \mathcal{V}$;\\
    Let the proxy node level Structural Data be $\hat d_v := d_v, v \in \mathcal{V}$, where $d_v$ denotes the degree of $v$;\\
    \For{$e=(u,v,+) \in \xi$}{
    $\mathcal{D} := \mathcal{D} \cup \{d_u,d_v,d_u+1,d_v+1\}$\;
    $\delta_k(\hat d_u):=\delta_k(\hat d_u)-1$, $\delta_k(\hat d_u+1):=\delta_k(\hat d_u+1)+1$\;
    $\delta_k(\hat d_v):=\delta_k(\hat d_v)-1$, $\delta_k(\hat d_v+1):=\delta_k(\hat d_v+1)+1$\;
    $\hat d_u := \hat d_u + 1$, $\hat d_v := \hat d_v + 1$, $\delta(u) := \delta(u) + 1$, $\delta(v) := \delta(v) + 1$\;
    \If{$\hat \alpha(u)$ == \NULL}{
    $\hat\alpha(u) := \hat \alpha(v)$\;
    $J_{n-c} := J_{n-c} \cup \{(u, \hat \alpha(v))\}$, $J_{c-n} := J_{c-n} \cup \{(\hat \alpha(v), u,+)\}$\;
    }
    \If{$\hat\alpha(v)$ == \NULL}{
    $\hat\alpha(v) := \hat \alpha(u)$\;
    $J_{n-c} := J_{n-c} \cup \{(v, \hat \alpha(u))\}$, $J_{c-n} := J_{c-n} \cup \{(\hat \alpha(u), v,+)\}$\;
    }
    $\mathcal{A} := \mathcal{A} \cup \{\hat\alpha(u), \hat\alpha(v)\}$\;
    \If{$\hat\alpha(v) == \hat\alpha(u)$}{
      $\delta_V(\hat\alpha(v)):= \delta_V(\hat\alpha(v)) + 2$\;
    }
    \If{$\hat\alpha(v) \neq \hat\alpha(u)$}{
      $\delta_V(\hat\alpha(v)) := \delta_V(\hat\alpha(v)) + 1$, $\delta_V(\hat\alpha(u)) := \delta_V(\hat\alpha(u)) + 1$\;
      $\delta_g(\hat\alpha(v)) := \delta_g(\hat\alpha(v)) + 1$, $\delta_g(\hat\alpha(u)) := \delta_g(\hat\alpha(u)) + 1$\;
    }
  }
  \Return the Adjustment from $G$ to $G'$.\\
  \caption{\fontsize{11}{12}\selectfont Naive adjustment generation algorithm (NAGA)}
  \label{algo:NAGA}
\end{algorithm}

\begin{algorithm}[]
    \fontsize{11}{12}\selectfont
  \SetAlgoLined
  
  \SetKwInOut{Input}{Input}\SetKwInOut{Output}{Output}
  \SetKwFunction{GetLength}{GetLength}
  \SetKwFunction{NULL}{NULL}
  \SetKwFunction{Partition}{Partition}
  \Input{The Structural Data ($d_i$, $V_\alpha$, $g_\alpha$, $m$, $\alpha(v_i)$, and $T_\alpha$) of the original graph $G$, an incremental sequence $\xi$ from $G$ to $G'$, and the iteration number $N$.}
  \Output{The Adjustment ($\delta(v_i)$, $\delta_k(d)$, $\delta_V(\alpha)$,  $\delta_g(\alpha)$, $n$, $J_{n-c}$, and $J_{c-n}$) from $G$ to $G'$.}
    $n := \GetLength(\xi)$, $J_{n-c}:=\emptyset$, $J_{c-n}:=\emptyset$\ $\delta(v_i):=0$, $\delta_k(d):=0$, $\delta_V(\alpha):=0$,  $\delta_g(\alpha):=0$;\\
    Let the proxy maps be $\hat \alpha(v_i) := \alpha(v_i), \hat{\hat{\alpha}}(v_i) := \alpha(v_i), v_i \in \mathcal{V}$;\\
    Let the proxy node-level Structural Data be $\hat d_v := d_v, v \in \mathcal{V}$, where $d_v$ denotes the degree of $v$;\\
    Let the involved node set be $I := \emptyset$;\\
    \tcp{Initialize the Adjustment}
    \For{$e=(u,v,op) \in \xi$}{
    $I:= I\cup \{u,v\}$;\\
    \uIf{$op == +$}{
        \uIf{$u, v$ are both existing nodes in $\mathcal{V}$}{
            Update the node-level Adjustment (using the proxy node-level Structural Data, the same as below);\\
            \uIf{$\alpha(u)$ and $\alpha(v)$ are both not None}{
            Update the community-level Adjustment without changing the community partitioning;\\
            }\uElseIf{$\alpha(u)$ or $\alpha(v)$ is None (suppose $\alpha(u)==$ None)}{
            $\delta_V(\alpha(u)) := \delta_V(\alpha(u)) + 1$; $\delta_g(\alpha(u)) := \delta_g(\alpha(u)) + 1$;\\
            }
        }\uElseIf{$u$ or $v$ does not exist (suppose $u$ does not exist)}{
            $\alpha(u) := \text{None}$;\\
            Update the node-level Adjustment;\\
            $\delta_V(\alpha(u)) := \delta_V(\alpha(u)) + 1$;
            $\delta_g(\alpha(u)) := \delta_g(\alpha(u)) + 1$;\\
            $J_{n-c} := J_{n-c} \cup \{(u, \text{None})\}$; 
            $J_{c-n} := J_{c-n} \cup \{(\text{None}, u, +)\}$;\\
        }\uElseIf{$u$ and $v$ are both not existing}{
            $\alpha(u) := \text{None}$, $\alpha(v) := \text{None}$;\\
            Update the node-level Adjustment;\\
            $J_{n-c} := J_{n-c} \cup \{(u, \text{None})\}$; $J_{n-c} := J_{n-c} \cup \{(v, \text{None})\}$;\\
            $J_{c-n} := J_{c-n} \cup \{(\text{None}, u, +)\}$; $J_{c-n} := J_{c-n} \cup \{(\text{None}, u, +)\}$;\\
        }
        Update the proxy node-level Strucutual Data as if the edge $e$ is added into the graph;\\
    }\ElseIf{$op == -$}{
        Update the node-level and the community-level Adjustment without changing the community partitioning;\\
        Update the proxy node-level Strucutual Data as if the edge $e$ is removed from the graph;\\
        }
    }
    \tcp{See the rest on the next page}
    \caption{\small{Node-shifting adjustment generation algorithm (NSGA)}}
    \label{algo:NSGA}
\end{algorithm}

\begin{algorithm}[t]
    \fontsize{11}{12}\selectfont   
    \LinesNumbered
    \setcounter{AlgoLine}{31}
    \tcp{The rest of Algorithm~\ref{algo:NSGA}}
    \tcp{Start node-shifting}
    $\tau := 1$;\\
    \While{$\tau \le N$ and $I \neq \emptyset$}{
    $\tilde I := \emptyset$, $X := \emptyset$;\\
    \For{each node $v \in I$}{
        Determine the OPC of $v$ denoted by $\alpha^*$ using $\hat \alpha(v)$;\\
        $X := X \cup \{(v, \alpha^*)\}$;\\
        \If{$\hat \alpha(v) \neq \alpha^*$}{
            Update the Adjustment as if $v$ moves into $T_{\alpha^*}$ using $\hat{\hat{\alpha}}(v)$;\\
            \For{each node $z \in$ Neighbor($v$)}{
                \If{$\hat \alpha(z) \neq \alpha^*$}{
                    $\tilde I := \tilde I \cup \{z\}$;\\
                }
            }
            $\hat{\hat{\alpha}}(v) := \alpha^* $;\\
        }
    }
    \For{each $(v, \alpha) \in X$}{
        $\hat \alpha(v) := \alpha $;\\
    }
    $I := \tilde I$, $\tau := \tau + 1$;\\
    
    }
  
  \Return the Adjustment from $G$ to $G'$.\\
\end{algorithm}

After getting the Adjustment, the updated two-dimensional structural entropy of $G'$ can then be incrementally calculated by:
\begin{equation}
H^{\mathcal{T}'}(G')=-\frac{1}{2m+2n}[\hat S'_N + \hat S'_C + \hat S'_G\log(2m+2n)],
\label{eq:incre}
\end{equation}
where $\hat S'_N$, $\hat S'_C$, and $\hat S'_G$ denote the incrementally updated Structural Expressions:
\begin{equation}
\begin{aligned}
\hat S'_N &= \hat S_N + \sum_{d \in \mathcal{D}} \delta_k (d) d\log d;\\
\hat S'_C &= \hat S_C + \sum_{\alpha \in \mathcal{A}} [(g_\alpha + \delta_g(\alpha) + V_\alpha + \delta_V(\alpha))\log (V_\alpha + \delta_V(\alpha)) - (g_\alpha + V_\alpha)\log V_\alpha]; \\
\hat S'_G &= \hat S_G + \sum_{\alpha \in \mathcal{A}} -\delta_g(\alpha).\\
\end{aligned}
\end{equation}

To implement the above incremental calculation process, we provide \textit{the Adjustment-based incremental updating algorithm (AIUA)} (Fig.~\ref{outline}\circled{2}).
Given the input, i.e., the Structural Data and Structural Expressions of the original graph and an Adjustment to the updated graph, we can compute the updated two-dimensional structural entropy incrementally, and update the Structural Data and Structural Expressions efficiently preparing for the next \textit{AIUA} process when a new Adjustment comes.
The pseudo-code of \textit{AIUA} is shown in Algorithm~\ref{algo:AIUA}.
The time complexity of updating the Structural Data is $O(|\phi_\lambda|+|\mathcal{A}|+|J_{n-c}|+|J_{c-n}|) \le O(n)$.
The time complexity of updating the Structural Expressions is $O(|\mathcal{D}|+|\mathcal{A}|) \le O(n)$.
The time complexity of calculating the updated two-dimensional structural entropy is $O(1)$.
In summary, the total time complexity of \textit{AIUA} is $O(n)$.

\begin{algorithm}[]
\fontsize{11}{12}\selectfont
  \SetAlgoLined
  \SetKwInOut{Input}{Input}\SetKwInOut{Output}{Output}
  \SetKwFunction{GetLength}{GetLength}
  \SetKwFunction{NULL}{NULL}
  \SetKwFunction{Partition}{Partition}
  \Input{The Structural Data ($d_i$, $V_\alpha$, $g_\alpha$, $m$, $\alpha(v_i)$, and $T_\alpha$) and the Structural Expressions ($\hat S_N$, $\hat S_C$, and $\hat S_G$) of the original graph $G$, and the Adjustment ($\delta(v_i)$, $\delta_k(d)$, $\delta_V(\alpha)$,  $\delta_g(\alpha)$, $n$, $J_{n-c}$, and $J_{c-n}$) from $G$ to $G'$.}
  \Output{The updated two-dimensional structural entropy $H^{\mathcal{T}'}(G')$, the updated Structural Data ($d'_i$, $V'_\alpha$, $g'_\alpha$, $m'$, $\alpha'(v_i)$, and $T'_\alpha$), and the updated Structural Expressions ($\hat S'_N$, $\hat S'_C$, and $\hat S'_G$).}
  \tcp{Update the Structural Data}
  \For{each $v_i \in \phi_\lambda$}{
  $d'_i := d_i + \delta(v_i)$;\\
  }
  
  \For{each $\alpha \in \mathcal{A}$}{
   $V'_\alpha := V_\alpha + \delta_V(\alpha)$; $g'_\alpha := g_\alpha + \delta_g(\alpha)$;\\
  }
  $m' = m + n$;\\
  \For{each $(v, \alpha) \in J_{n-c}$}{
  $\alpha'(v):=\alpha$;\\
  }
  $T'_\alpha := T_\alpha$ for all $\alpha \in A$;\\
  \For{each $(\alpha, v, op) \in J_{c-n}$}{
  \If{$op == +$}{
      $T'_{\alpha} := T'_{\alpha}\cup\{v\}$\;
    }
  \If{$op == -$}{
      $T'_{\alpha} := T'_{\alpha}/\{v\}$\;
    }
  }
  \tcp{Update the Structural Expressions}
  $\hat S'_N := \hat S_N$; $\hat S'_C := \hat S_C$; $\hat S'_G := \hat S_G$;\\
  \For{each $d \in \mathcal{D}$}{
   $S'_N := S'_N  + \delta_k(d) d \log d$;\\
  }
  \For{each $\alpha \in \mathcal{A}$}{
   $\hat S'_C := \hat S'_C + (g_\alpha + \delta_g(\alpha) + V_\alpha + \delta_V(\alpha))\log (V_\alpha + \delta_V(\alpha)) - (g_\alpha + V_\alpha)\log V_\alpha$;\\
   $\hat S'_G := \hat S'_G - \delta_g(\alpha)$;\\
  }
  \tcp{Calculate the updated two-dimensional structural entropy}
  $H^{\mathcal{T}'}(G') := -\frac{1}{2m+2n}[\hat S'_N + \hat S'_C + \hat S'_G\log(2m+2n)]$;\\
  
  \Return $H^{\mathcal{T}'}(G')$, the updated Structural Data, and the updated Structural Expressions.\\
  \caption{\fontsize{10}{12}\selectfont Adjustment-based incremental updating algorithm (AIUA)}
  \label{algo:AIUA}
\end{algorithm}

\subsection{
\revised{Baseline: the Traditional Offline Algorithm}
}
\textit{The traditional offline algorithm (TOA)} reconstructs the encoding tree for each updated graph and calculates the updated two-dimensional structural entropy by definition.
\textit{TOA} consists of the following four steps.
Firstly, it generates the updated graph by combining the original graph and the incremental sequence~(\textcircled{a} in Fig.~\ref{outline}).
Secondly, it partitions the graph node set into communities using several different static community detection algorithms, e.g., Infomap~\citep{rosvall2008maps}, Louvain~\citep{louvain}, and Leiden~\citep{traag2019louvain}, to construct the two-dimensional encoding tree~(\textcircled{b} in Fig.~\ref{outline}).
Thirdly, the node-level, community-level, and graph-level Structural Data of the updated graph is counted and saved~(\textcircled{c} in Fig.~\ref{outline}).
Finally, the updated structural entropy is computed via Eq.~(\ref{2dSE-updated})~(\textcircled{d} in Fig.~\ref{outline}).
The total time cost of \textit{TOA} is $O(|\mathcal{E}|+n)$ plus the cost of the chosen community detection algorithm.
The pseudo-code of \textit{TOA} is shown in Algorithm~\ref{algo:TOA}.

\begin{algorithm}[]
\fontsize{11}{12}\selectfont
  \SetAlgoLined
  \SetKwInOut{Input}{Input}\SetKwInOut{Output}{Output}
  \SetKwFunction{CMB}{CMB}
  \Input{The original graph $G$ and an incremental sequence $\xi$.}
  \Output{The updated two-dimensional structural entropy $H^{\mathcal{T}'}(G')$.}
  Get the updated graph $G'$ by combining $G$ and $\xi$\;
  Construct the two-dimensional encoding tree $\mathcal{T}'$.\;
  Get the degree $d'_i$ of each node $v_i \in \mathcal{V}'$\;
  Get the volume $V'_{\alpha}$ and the cut edge number $g'_{\alpha}$ of $\alpha \in A'$\;
  Get the total edge number $m'$ of $G'$\;
  Obtain $H^{\mathcal{T}'}(G')$ via Eq.~(\ref{2dSE-updated})\;
  \Return $H^{\mathcal{T}'}(G')$\;
  \caption{Traditional Offline Algorithm (TOA)}
  \label{algo:TOA}
\end{algorithm}

\section{\revised{Extensions on More Complex Graphs}}\label{sec:ext}
\revised{
In this section, we discuss the feasibility of the extension to weighted or directed graphs of our methods.
First, we argue that the method for undirected weighted graphs can be extended naturally from that of undirected unweighted graphs.
Second, we analyze the fundamental differences between the paradigm of structural entropy incremental computation for directed graphs and that for undirected graphs and present new methods for calculating one-dimensional structural entropy incrementally on directed weighted graphs.
}
\subsection{\revised{Undirected Weighted Graphs}}
\revised{
The incremental measurement method of structural entropy for undirected weighted graphs can be intuitively and easily extended from the methods for undirected unweighted graphs proposed earlier.
In the following, we first introduce the definition of two-dimensional structural entropy of undirected weighted graphs.
Then we update the definition of the Adjustment and propose an incremental formula for structural entropy computation under the new circumstance.
}

\noindent
\revised{
\textbf{Two-dimensional structural entropy of undirected weighted graphs.}
An undirected weighted graph can be denoted as $G_W = (\mathcal{V}, \mathcal{E}, w)$, where $w(e) \in \mathbb{R}^+$ represents the weight of $e \in \mathcal{E}$.
For any $e \notin \mathcal{E}$, let $w(e) = 0$.
According to Li and Pan~\citep{structural-entropy}, the definition of the encoding tree for an undirected weighted graph remains unchanged, and the two-dimensional structural entropy of $G_W = (\mathcal{V}, \mathcal{E}, w)$ by its two-dimensional encoding tree $\mathcal{T}$ is defined as
}
\revised{
\begin{equation}
H^\mathcal{T}(G_W)=\sum_{\alpha_i \in A}(-\frac{g_{\alpha_i}}{V_\lambda}\log\frac{V_{\alpha_i}}{V_\lambda}
+\sum_{v_j\in T_{\alpha_i}}-\frac{d_{j}}{V_\lambda}\log\frac{d_j}{V_{\alpha_i}}),
\label{eq:2dSE-uwg}
\end{equation}}
\revised{where new degree $d_j = \sum_{v_i \in \mathcal{N}(v_j)} w(v_j, v_i)$ ($\mathcal{N}(v)$ denotes the set of neighbors of node $v$), new volume $V_{\alpha} = \sum_{v_i \in T_\alpha} d_i$, and new cut edge number $g_{\alpha}$ is replaced with the sum of the weights of the edges with exactly one endpoint in $T_\alpha$.
}

\noindent
\revised{
\textbf{Adjustment definition of undirected weighted graphs.}
Due to the continuity of edge weights, the original node level and graph level Adjustment definitions (Definition~\ref{def:adj}) no longer apply.
We renew the Adjustment definitions of the two levels as follows:
\begin{enumerate}
\item (node level) the total weight change $\delta(v_i) = d'_i - d_i$ for each node $v_i \in \phi_\lambda$;
\item (graph level) the total weight change of the whole graph denoted by $\delta_V(\lambda)$.
\end{enumerate}
}

\noindent\textbf{\revised{Incremental formula for structural entropy computation.}}
\revised{
According to Eq.~(\ref{eq:2dSE-uwg}), the incremental computation formula for undirected weighted graphs can then be formulated as (extended from Eq.~(\ref{eq:incre}))}
\begin{equation}
\revised{
H^{\mathcal{T}'}(G'_W)=-\frac{1}{V_\lambda+\delta_V(\lambda)}[\hat S'_N + \hat S'_C + \hat S'_G\log(V_\lambda+\delta_V(\lambda))],
}
\label{eq:incre-uwg}
\end{equation}
\revised{
where
}
\begin{equation}
\revised{
\begin{aligned}
\hat S'_N &= \hat S_N + \sum_{v_k \in \phi_\lambda}[(d_k+\delta(v_k))\log(d_k+\delta(v_k)) - d_k\log{d_k}];\\
\hat S'_C &= \hat S_C + \sum_{\alpha \in \mathcal{A}} [(g_\alpha + \delta_g(\alpha) + V_\alpha + \delta_V(\alpha))\log (V_\alpha + \delta_V(\alpha)) - (g_\alpha + V_\alpha)\log V_\alpha]; \\
\hat S'_G &= \hat S_G + \sum_{\alpha \in \mathcal{A}} -\delta_g(\alpha).\\
\end{aligned}
}
\end{equation}

\subsection{\revised{Directed Weighted Graphs}}
\revised{The main method proposed in this paper is difficult to transfer to directed graph scenarios since the measurement of the structural entropy of directed graphs is fundamentally different from that of undirected graphs.
The key difference is that the directed graph needs to be transferred into a transition matrix and stationary distribution needs computed.
In this part, we briefly present an incremental scheme for measuring the one-dimensional structural entropy of directed weighted graphs since the incremental computation of two-dimensional structural entropy is quite complex.
Specifically, we first define the directed weighted graph and its non-negative matrix representation.
After that, we introduce the formula of the structural entropy of directed weighted graphs~\citep{structural-entropy}.
Finally, we review the traditional methods, namely Eigenvector Calculation and Global Aggregation, for accurately or approximately calculating the one-dimensional structural entropy of directed weighted graphs
and then propose an incremental iterative approximation algorithm, i.e., Local Propagation.
}

\subsubsection{\revised{Directed Weighted Graph and Non-Negative Matrix}}
\begin{definition}[\revised{Directed Weighted Graph}]
\revised{
A directed weighted graph can be denoted as $G_{DW} = (\mathcal{V},\mathcal{E}, f)$, which satisfies the following properties:
\begin{enumerate}[(1)]
    \item $\mathcal{V} = \{v_1, v_2, ..., v_N\}$ is the node set;
    \item $\mathcal{E}$ denotes the set of directed edges $e = (v_i, v_j)$;
    \item for each directed edge $e = (v_i, v_j) \in \mathcal{E}$, $f(e) = f(v_i, v_j) >0$ is the edge weight from $v_i$ to $v_j$.
    For $e = (v_i, v_j) \notin \mathcal{E}$, we denote $f(e) = f(v_i, v_j) = 0$.
\end{enumerate}
}
\label{def:wdg}
\end{definition}

\revised{
The definition of directed weighted graphs is shown in Definition~\ref{def:wdg}.
Given a directed weighted graph $G_{DW} = (\mathcal{V},\mathcal{E}, f)$ with $N$ nodes, we fix the nodes in $\mathcal{V}$ in an order like $v_1, v_2, ..., v_N$.
For $i,j \in \{1,2,..,N\}$, we define $a_{ij} = f(v_i, v_j)$.
Then we can obtain a non-negative matrix $\mathbf{A} = (a_{ij})$, named as a matrix of $G_{DW}$.
In other words, a directed weighted graph can be represented by a non-negative matrix.
The definition of the non-negative matrix representation of directed weighted graphs is presented in Definition~\ref{def:matrix}.
}

\begin{definition}[\revised{Directed Weighted Graph Represented by Non-Negative Matrix}]
\revised{
A directed weighted graph $G_{DW} = (\mathcal{V},\mathcal{E}, f)$ can be denoted as an $N\times N$ non-negative matrix $\mathbf{A} = (a_{ij}) \in \mathbb{R}^{N\times N}_{\ge 0}$, where for each $i, j \in \{1,2,...,N\}$, $a_{ij} = f(v_i, v_j) \ge 0$.
}
\label{def:matrix}
\end{definition}

\subsubsection{\revised{One-Dimensional Structural Entropy of Directed Weighted Graphs}}

\revised{
For direct graphs, the structural entropy is an uncertainty measurement defined on strongly connected graphs whose matrices must be irreducible (Definition~\ref{def:im}).
}

\begin{definition}[\revised{Irreducible Matrix}]
\revised{
Given a non-negative matrix $\mathbf{A}\in \mathbb{R}^{N\times N}_{\ge 0}$, if there exists a permutation matrix $\mathbf{P}$ such that 
\begin{equation}
    \mathbf{P} \mathbf{A} \mathbf{P} = 
    \begin{bmatrix}
        \mathbf{X} & \mathbf{Y} \\
        \mathbf{0} & \mathbf{Z}
    \end{bmatrix},
\end{equation}
where $\mathbf{X}$ and $\mathbf{Z}$ are square matrices, then $\mathbf{A}$ is said to be a reducible matrix, otherwise $\mathbf{A}$ is an irreducible matrix.
}
\label{def:im}
\end{definition}

\revised{
Normalize each row in an irreducible matrix $\mathbf{A}$, i.e., for each $i$ and $j$, let
\begin{equation}
    b_{ij} = \frac{a_{ij}}{\sum_{k = 1}^N a_{ik}}.
\end{equation}
Then we can get a normalized irreducible matrix $\mathbf{B} = (b_{ij})$ where the sum of elements in each row is $1$.
$b_{ij}$ is called the normalized edge weight.
There is a fundamental theorem for normalized irreducible matrices like $\mathbf{B}$.
}

\begin{theorem}[\revised{Perron Forbenius}]
\revised{
If $\mathbf{B}\in \mathbb{R}^{N\times N}_{\ge 0}$ is an irreducible matrix where for each $i$, $\sum_{k=1}^N b_{ik}= 1$, then
\begin{enumerate}[(1)]
    \item the maximum eigenvalue of $\mathbf{B}$ is $1$;
    \item the maximum eigenvalue of $\mathbf{B}$ has a unique left eigenvector;
    \item The unique left eigenvector of the maximum eigenvalue of $\mathbf{B}$ is a probability distribution, denoted by $\pi = [\pi_1, \pi_2, ..., \pi_N]$.
\end{enumerate}
}
\label{the:pf}
\end{theorem}

\revised{
According to Theorem~\ref{the:pf}, the stationary distribution of $\mathbf{A}$ can then be defined in Definition~\ref{def:sd}.
}

\begin{definition}[\revised{Stationary Distribution}]
\revised{
Let $\mathbf{A}\in \mathbb{R}^{N\times N}_{\ge 0}$ be an irreducible matrix and $\mathbf{B}$ represents the normalized matrix of $\mathbf{A}$ where for each $i$, $\sum_{k=1}^N b_{ik}$ $= 1$.
The stationary distribution of $\mathbf{A}$ is defined as the unique left eigenvector of the maximum eigenvalue of $\mathbf{B}$, denoted by $\pi = [\pi_1, \pi_2, ..., \pi_N]$.
}
\label{def:sta}
\end{definition}

\revised{
Finally, the one-dimensional structural entropy (Definition~\ref{def:1dSE}) of an irreducible non-negative matrix (or a directed weighted graph) is defined as the Shannon entropy of the stationary distribution, which measures the total amount of uncertainty embedded in a directed weighted graph.
}

\begin{definition}[\revised{One-Dimensional Structural Entropy}]
\revised{
Let $\mathbf{A}\in \mathbb{R}^{N\times N}_{\ge 0}$ be an irreducible matrix and $\pi = [\pi_1, \pi_2, ..., \pi_N]$ is the stationary distribution of $\mathbf{A}$.
The one-dimensional structural entropy is defined as
}
\revised{
\begin{equation}
\mathcal{H}^1(\mathbf{A}) = -\sum_{i=1}^N \pi_i \log_2 \pi_i.
\label{eq:1dSE}
\end{equation}
}
\label{def:1dSE}
\end{definition}

\subsubsection{\revised{Incremental Measurement of One-Dimensional Structural Entropy}}

\revised{
Generally, the exact value of one-dimensional structural entropy can be obtained by Eq.~(\ref{eq:1dSE}) in $O(n)$ given the stationary distribution.
However, calculating the exact stationary distribution needs to solve the eigenvectors of the normalized irreducible matrix, which usually costs $O(n^3)$.
In dynamic scenarios, the directed weighted graph evolves as time passes by.
When an irreducible matrix $\mathbf{A}\in \mathbb{R}^{N\times N}_{\ge 0}$ representing a directed weighted graph gets an incremental $\Delta \mathbf{A}$, it becomes an updated irreducible matrix $\mathbf{A}' = \mathbf{A} + \Delta \mathbf{A}$.
Let $n = ||\Delta \mathbf{A}||_0$ denote the incremental size.
Let $\mathbf{B}$ and $\mathbf{B}'$ be the normalized matrices of $\mathbf{A}$ and $\mathbf{A}'$.
Define $\Delta \mathbf{B} = \mathbf{B}' - \mathbf{B}$.
We can calculate $\Delta \mathbf{B} = (\Delta b_{ij})$ from $\mathbf{A}$ and $\Delta \mathbf{A}$ by
\begin{equation}
\Delta b_{ij} = b'_{ij} - b_{ij} = \frac{a'_{ij}}{\sum_{k=1}^{N} a'_{ik}} - \frac{a_{ij}}{\sum_{k=1}^{N} a_{ik}} = \frac{a_{ij} + \Delta a_{ij}}{(\sum_{k=1}^{N} a_{ik} + \Delta a_{ik})} - \frac{a_{ij}}{\sum_{k=1}^{N} a_{ik}}.
\end{equation}
Since a non-zero element of $\Delta \mathbf{A}$ may influence at most all $N$ elements of a row in $\mathbf{B}$, $||\Delta \mathbf{B}||_0$, named as the normalized incremental size, will be no more than $nN$.
In addition, the number of the influenced rows in $\mathbf{B}$, denoted by $n_B$, will be no more than $n$.
In the following, three ways are listed to compute the updated one-dimensional structural entropy.
}

\noindent\textbf{\revised{Exact value calculation by Eigenvector Calculation.}}
\revised{
The first way is to calculate the exact value of the updated one-dimensional structural entropy by Definition~\ref{def:sta} and~\ref{def:1dSE}.
Specifically, the new exact stationary distribution $\pi'$ is first computed by solving the left eigenvector of the maximum eigenvalue of $\mathbf{B}'$.
Then the updated one-dimensional structural entropy can be calculated by
\begin{equation}
\mathcal{H}^1(\mathbf{A'}) = -\sum_{i=1}^N \pi'_i \log_2 \pi'_i.
\label{eq:updated-1dse}
\end{equation}
Generally, the total time complexity of this way is $O(N^3)$.
}

\noindent\textbf{\revised{Approximate value calculation by Global Aggregation.}}
\revised{
According to the theory of the Markov chain, the approximate stationary distribution can be approximated by iteratively right-multiplying $\mathbf{B}'$, i.e.,
\begin{equation}
\pi^{(\theta)} = \pi \mathbf{B}'^\theta = \pi (\mathbf{B} +\Delta \mathbf{B})^\theta,
\label{eq:theta-pi}
\end{equation}
where $\theta \in \mathbb{N}$ is the iteration number.
Whenever $\pi^{(\theta)}$ updates to $\pi^{(\theta+1)}$ by right-multiplying $\mathbf{B} +\Delta \mathbf{B}$, the updating process is equivalent to an information aggregation operation on graphs along directed edges, i.e., for each node $v_i$, 
\begin{equation}
    \pi^{(\theta + 1)}_i = \sum_{v_j \in \mathcal{N}(v_i)} \pi^{(\theta)}_j b_{ji},
    \label{eq:theta-pi-graph}
\end{equation}
where $\mathcal{N}(v_i)$ denotes the neighbors of $v_i$.
Since for each iteration, all nodes need to be updated, we name this calculation method as ``Global Aggregation''.
Finally, we compute the Shannon entropy of $\pi^{(\theta)}$ as the approximate one-dimensional structural entropy:
\begin{equation}
\mathcal{H}^1(\mathbf{A'}) \approx -\sum_{i=1}^N \pi^{(\theta)}_i \log_2 \pi^{(\theta)}_i.
\end{equation}
The total computational complexity of Global Aggregation is about $O(\theta N^2)$ in the form of matrix multiplication using Eq.~(\ref{eq:theta-pi}).
Utilizing graph perspective (Eq.~(\ref{eq:theta-pi-graph})), the time complexity can be reduced to $O(\theta N \overline{d}) = O(\theta |\mathcal{E}|)$, where $\overline{d}$ denotes the mean direct successor number of all graph nodes.
}

\noindent\textbf{\revised{Fast approximate value calculation by Local Propagation.}}
\revised{In Global Aggregation, all nodes and edges need to be traversed in each iteration, which leads to high computational redundancy.
In this part, we propose a new method for fast approximation of the updated one-dimensional structural entropy, namely Local Propagation.
As the name suggests, the key idea is to use an information propagation scheme involving only changed local nodes to further reduce the redundancy of the aggregation process in Eq.~(\ref{eq:theta-pi-graph}).
}

\begin{figure}
    \centering
    \includegraphics[width = 0.9\linewidth]{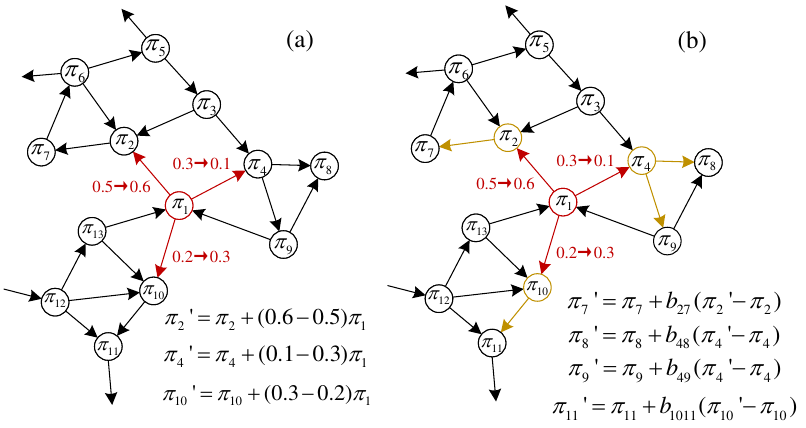}
    \caption{
    \revised{
    An illustration of Local Propagation.
    Note that all edge weights are normalized.
    In the first step (a), the red arrows denote the edges whose weight changes after getting an incremental, i.e., the involved edges.
    The stationary distributions of the involved nodes ($\pi_2, \pi_4, \pi_{10}$) are updated.
    In the second step (b), the direct successors of the last involved nodes update their stationary distributions and become new involved nodes.
    Then the procedure of (b) is repeated until the maximum iteration number is reached.
    }
    }
    \label{fig:lp}
\end{figure}

\revised{
In particular, Local Propagation contains two steps.
In the first step (Fig.~\ref{fig:lp}(a)), we define the set of directed edges in $\mathbf{B}$ influenced by $\Delta \mathbf{B}$ as ``involved edges'' (red edges in Fig.~\ref{fig:lp}(a)).
We then let $\pi^{(1)} = \pi^{(0)}$.
For each involved edge $(v_i, v_j)$, the stationary distribution value of the pointed node $v_j$ is updated by
\begin{equation}
    \pi^{(1)}_j \leftarrow \pi^{(1)}_j + \Delta b_{ij} \pi_i,
\end{equation}
and $v_j$ is added into an ``involved nodes'' set denoted by $I^{(1)}$.
The time complexity of the first step is linearly correlated to the size of the involved edge set, i.e., $O(nN)$.
}
\revised{
In the second step, we repeat the following procedure for $\theta - 1$ times.
Let $\pi^{(\theta+1)} = \pi^{(\theta)}$.
For each node $v_i$ in $I^{(\theta)}$, we update the stationary distribution values of all $v_i$'s direct successors (like Fig.~\ref{fig:lp}(b)) by
\begin{equation}
    \pi^{(\theta+1)}_j \leftarrow \pi^{(\theta+1)}_j + b'_{ij}(\pi^{(\theta)}_i-\pi^{(\theta-1)}_i), v_j \in \mathcal{N}_s(v_i),
\end{equation}
where $\mathcal{N}_s(v_i)$ denotes the direct successors of $v_i$.
After all nodes in $I^{(\theta)}$ are traversed, $I^{(\theta)}$ is updated as $I^{(\theta+1)}$, i.e., all the direct successors of nodes of the original set $I^{(\theta)}$.
For each iteration, the time complexity is $O(|I^{(\theta)}|\overline{d}^{(\theta)})$, where $\overline{d}^{(\theta)}$ denotes the mean direct successor number of nodes in $I^{(\theta)}$.
In other words, let $E^{(\theta)}$ ($\theta \ge 1$) denote the set of edges whose starting points belong to $I^{(\theta)}$, we have $|I^{(\theta)}|\overline{d}^{(\theta)} = |E^{(\theta)}|$.
Therefore, the time complexity of each iteration is $O(|E^{(\theta)}|)$ which must be less or equal to the complexity of Eq.~(\ref{eq:theta-pi-graph}), $O(|\mathcal{E}|)$.
}


\section{Experiments and Evaluations}\label{sec:eval}
In this section, we conduct extensive experiments based on the application of dynamic graph real-time monitoring and community optimization.
Below we first describe the $3$ artificial dynamic graph datasets and $3$ real-world datasets.
Then we give the experimental results and analysis.

\subsection{Artificial Datasets}\label{sec:ad}
First, we generate $3$ different initial states of the dynamic graphs by utilizing the \textit{random\_partition\_graph} (Random) \citep{2009Community}, \textit{gaussian\_random\_partition} \textit{\_graph} (Gaussian) \citep{brandes2003experiments}, and \textit{stochastic\_block\_model} (SBM) \citep{holland1983stochastic} methods in ``Networkx"~\citep{networkx} (a Python library).
\revised{Parameter descriptions of the three methods are listed below:}
\begin{itemize}
\item \revised{\textbf{Random parameters:}}
This method has $3$ parameters.
The first parameter is a list of community sizes $\mathcal{S} = [s_1, s_2,...]$, which denotes the node number of each community of the initial state.
The other two parameters are two probabilities $p_\mathrm{in}$ and $p_\mathrm{ac}$.
Nodes in the same community are connected with $p_\mathrm{in}$ and nodes across different communities are connected with $p_\mathrm{ac}$.
\item \revised{\textbf{Gaussian parameters:}}
\revised{
This method creates $k$ communities each with a size drawn from a normal distribution with mean $s$ and variance $s/v$. 
Nodes are connected within communities with probability $p_\mathrm{in}$ and between communities with probability $p_\mathrm{ac}$.
}
\item \revised{\textbf{SBM parameters:}}
\revised{
This method partitions the nodes in communities of given sizes $\mathcal{S} = [s_1, s_2,...]$, and places edges between pairs of nodes independently, with a probability that depends on the communities.
}
\end{itemize}

\begin{figure}[!b]
    \centering
    \includegraphics[width = \linewidth]{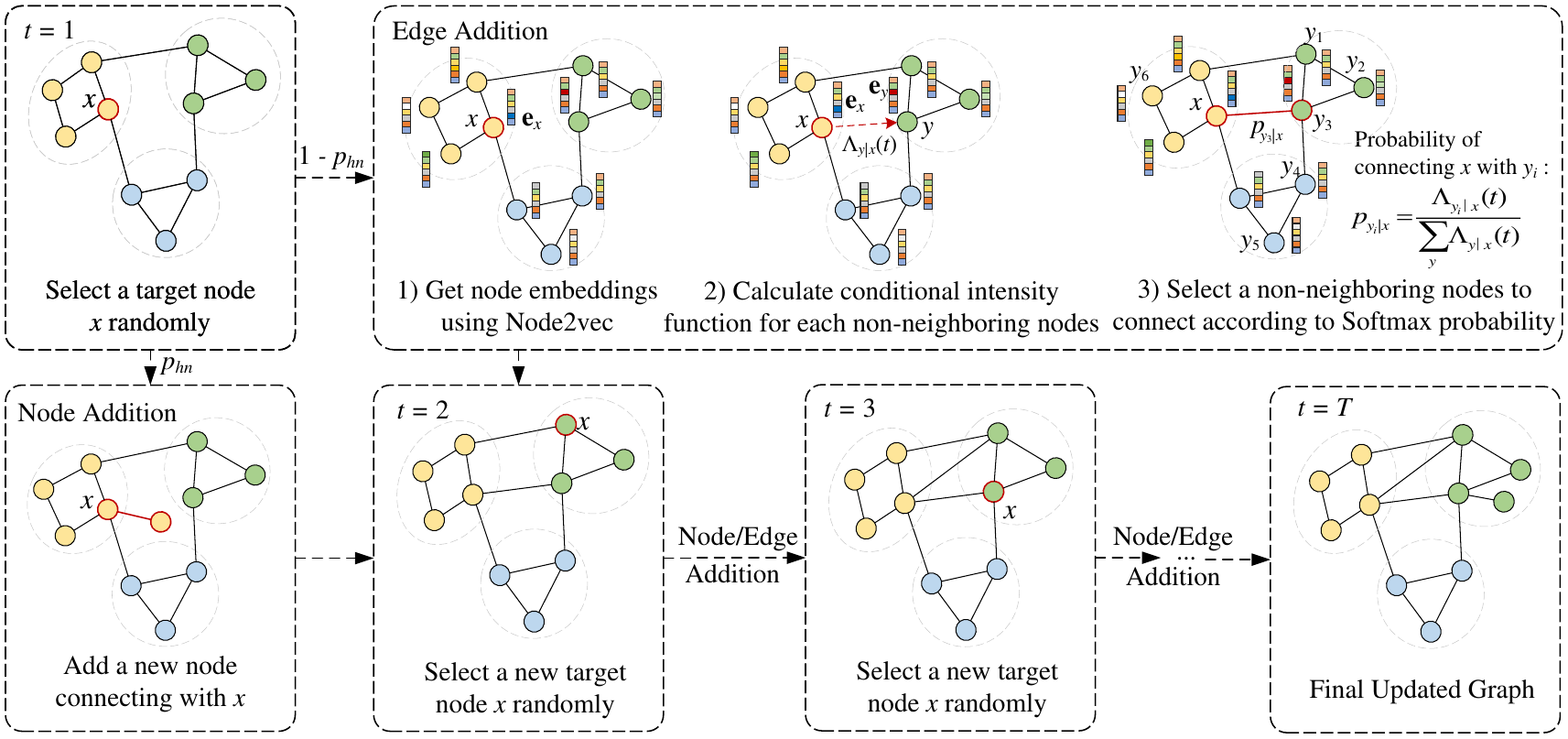}
    \caption{
    \revised{The generation process of the artificial Hawkes datasets.}
    }
    \label{fig:hawkes}
\end{figure}

After that, we generate incremental sequences and updated graphs for each initial state by Hawkes Process~\citep{hawkes-origin} referring to some settings of Zuo et al.~\citep{hawkes-zuo}.
Hawkes Process~\citep{hawkes-origin} models discrete sequence events by assuming that historical events can influence the occurrence of current events. 
\revised{
In this process, we first randomly choose a node $x$ to be the target node (Fig.~\ref{fig:hawkes}($t=1$)).
Second, we add edges or nodes with given probabilities.
Specifically, with probability $p_\mathrm{hn}$ (Fig.~\ref{fig:hawkes}(Node Addition)), we connect a new node with $x$.
With probability $1-p_\mathrm{hn}$ (Fig.~\ref{fig:hawkes}(Edge Addition)), we (1) use Node2vec~\citep{node2vec} to get embedding vectors of all nodes; (2) calculate the conditional intensity function $\Lambda_{y_i|x}$ between $x$ and each of its non-neighboring nodes $y$:
}
\begin{equation}
    \Lambda_{y \mid x}(t)=-||\textbf{e}_x-\textbf{e}_y||^2 + \sum_{t_{h}<t} -||\textbf{e}_h-\textbf{e}_y||^2\exp \left(-\delta_x\left(t-t_{h}\right)\right),
\end{equation}
\revised{
where $\textbf{e}_x,\textbf{e}_y$ refers to the embedding vectors of nodes $x$ and $y$,
$h$ refers to the historical neighbor nodes connected to $x$ at time $t_h$ before $t$,
and $\delta_x$ refers to the discount rate, defined in this paper as the number of neighbors of $x$;
and (3) add an edge between $x$ and $y_i$ with the Softmax conditional probability:
}
\begin{equation}
    p_{y_i|x} = \frac{\Lambda_{y_i|x}}{\sum_{y}\Lambda_{y|x}}.
\end{equation}
\revised{
Subsequently, we repeat the above two steps to generate incremental sequences and updated graphs (Fig.~\ref{fig:hawkes}($t=2, 3,..., T$)).
The chosen parameter settings of the initial states and Hawkes Process are described in Table~\ref{tab:para}.
}

\begin{table}[t]
\centering
\caption{
\revised{
Parameter Values of the Generated Initial States and Hawkes Process.
}
}
\resizebox{\columnwidth}{!}{%
\begin{threeparttable}

\begin{tabular}{@{}ll@{}}
\toprule
                       & Parameter values \\ \midrule
Random initial state   & $\mathcal{S} = [800, 1000, 1200, 1400, 1600]$, $p_\mathrm{in}=0.05$, and $p_\mathrm{ac}=0.001$.  \\
Gaussian initial state & The total node number is $6000$, $s = 1000$, $v = 100$, $p_\mathrm{in}=0.05$, and $p_\mathrm{ac}=0.001$.        \\
SBM initial state      & $\mathcal{S} = [800, 1000, 1200, 1400, 1600]$, $p_\mathrm{in} \sim $\tnote{*} $ \mathcal{U}(0.002, 0.01)$, and $p_\mathrm{ac} \sim \mathcal{U}(0.01, 0.05)$.  \\
\midrule
Hawkes process         & The dimension of the embedding vectors is set as $16$ and $p_\mathrm{hn}$ is $0.05$.                \\ \bottomrule
\end{tabular}%

\begin{tablenotes}
    \item[*]$\mathcal{U}$ denotes uniform distribution.
\end{tablenotes}
    
\end{threeparttable}

}
\label{tab:para}
\end{table}



\begin{table}[t]
\centering
\caption{
\revised{Statistics Description of the Artificial and Real-World Datasets.}
}
\resizebox{0.78\columnwidth}{!}{%
\begin{tabular}{@{}clcccc@{}}
\toprule
\revised{Datasets} & \multicolumn{1}{c}{$|\mathcal{V}_0|$} & $|\mathcal{E}_0|$ & $\mathbb{E}(|\Delta\mathcal{V}|)$ & $\mathbb{E}(|\Delta\mathcal{E}|)$ & $\#$ of snapshots \\ \midrule
\revised{Random-Hawkes}   & \revised{6,000}  & \revised{203,659} & \revised{1,017} & \revised{20,365} & 20 \\
\revised{Gaussian-Hawkes} & \revised{6,000}  & \revised{164,482} & \revised{596} & \revised{11,942} & 20 \\
\revised{SBM-Hawkes}      & \revised{6,000}  & \revised{176,526} & \revised{592} & \revised{11,942} & 20 \\ \midrule
Cit-HepPh                 & \revised{25,656} & \revised{132,119} & \revised{235}   & \revised{10,727} & 20 \\
\revised{DBLP}            & \revised{7,184}  & \revised{13,451}  & \revised{3,379} & \revised{7,907}  & 20 \\
Facebook                  & \revised{14,094} & \revised{72,809}  & \revised{2,233} & \revised{26,000} & 20 \\ \bottomrule
\end{tabular}%
}
\label{tab:datasets}
\end{table}

\subsection{Real-World Datasets}
For the real-world datasets, we choose Cit-HepPh~\citep{leskovec2005graphs}, DBLP~\citep{bader2013graph}, and Facebook~\citep{viswanath2009evolution} to conduct our experiments.
Cit-HepPh is a citation network in the field of high-energy physics phenomenology from $1993$ to $2003$.
\revised{DBLP contains a co-authorship network of computer science papers from $1954$ to $2015$ in which authors are represented as vertices and co-authors are linked by an edge.}
Facebook records the establishment process of the user friendship relationship of about $52\%$ Facebook users in New Orleans from $2006$ to $2009$.
For each dataset, we cut out $21$ consecutive snapshots (an initial state and $20$ updated graphs).
Since structural entropy is only defined on connected graphs, we only preserve the largest connected component for each snapshot.
Overall, the statistics of the artificial and real-world datasets are briefly shown in Table~\ref{tab:datasets}.

\subsection{Results and Analysis}

\subsubsection{Application: Dynamic Graph Real-Time Monitoring and Community Optimization}

\label{sec:app}
In this application, we aim to optimize the community partitioning and monitor the corresponding two-dimensional structural entropy by our incremental algorithms, i.e., \textit{NAGA+AIUA} and \textit{NSGA+AIUA}, and the baseline \textit{TOA} to quantify the community quality in real-time for each snapshot of a dynamic graph.
Specifically, for each dataset, we first choose a static community detection method (referred to as static methods) from Infomap~\citep{rosvall2008maps}, Louvain~\citep{louvain}, and Leiden~\citep{traag2019louvain} to generate the initial state's community partitioning.
\newrevised{In this paper, Louvain is complemented by the \textit{louvain\_communities} method in ``Networkx''~\citep{networkx}.
Infomap and Leiden are complemented by the \textit{community\_infomap} and the \textit{community\_leiden} methods from the Python library ``igraph''~\citep{csardi2006igraph}, respectively.
Louvain and Infomap algorithms take default parameters while in Leiden we use ``Modularity'' as the objective instead of the original setting  ``Constant Potts Model (CPM)''.
This is because Leiden with CPM cannot effectively partition the communities on our datasets, as all partitions generated by Leiden with CPM contain only one node.}
Then we use \textit{NAGA+AIUA}, \textit{NSGA+AIUA}, and \textit{TOA} to respectively measure the updated two-dimensional structural entropy at each time stamp.
The \newrevised{default} maximum iteration number of \textit{NSGA} is set as $5$ \newrevised{and the reason is discussed in Section~\ref{sec:time}}.

\begin{figure}[h]
    \centering
    \includegraphics[width = \linewidth]{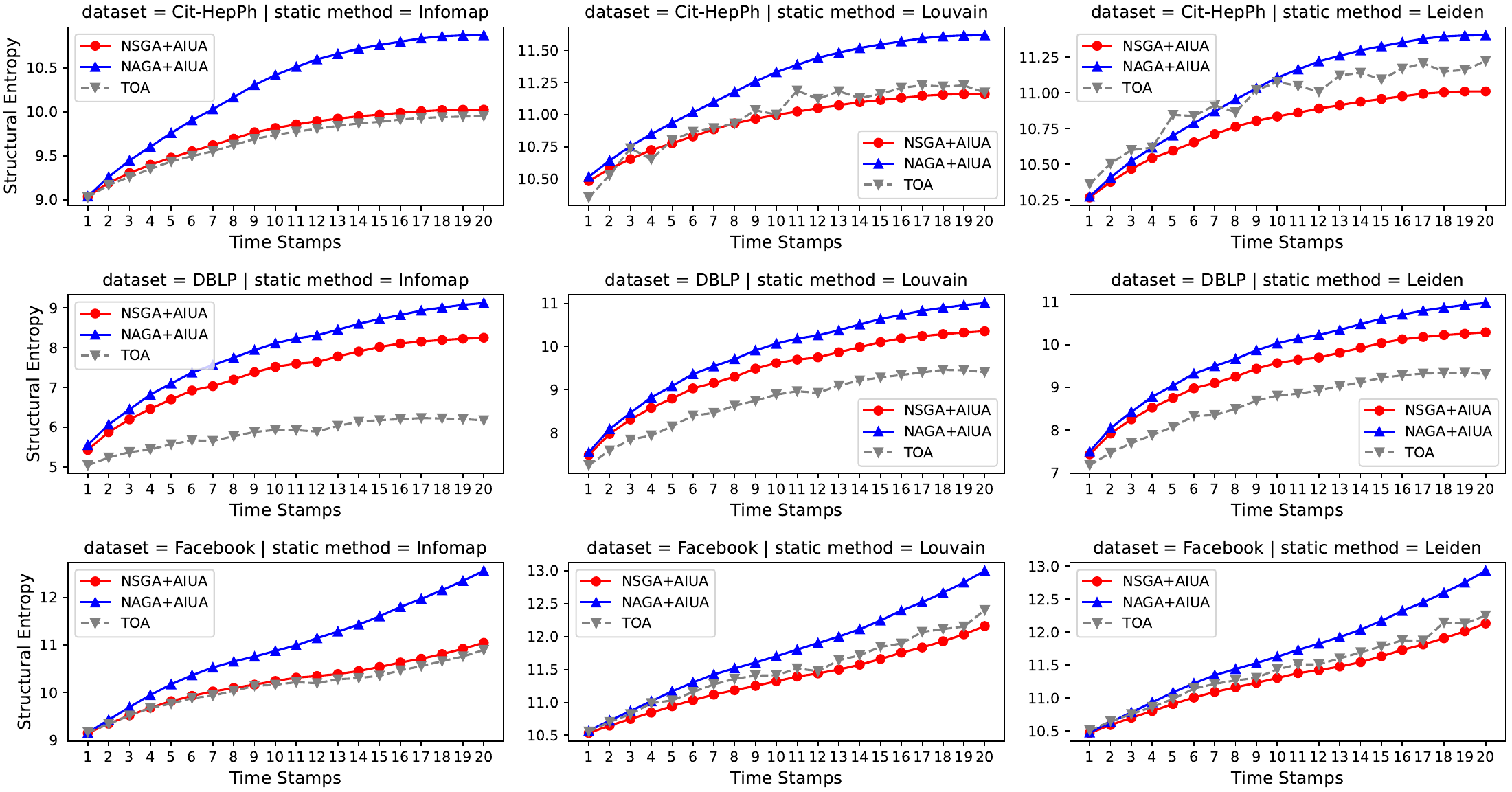}
    \caption{The updated structural entropy measured by \textit{NAGA+AIUA}, \textit{NSGA+AIUA}, and \textit{TOA} on real-world datasets with different static methods.
    Lower structural entropy represents better performance.
    }
    \label{fig:exp1}
\end{figure}

\begin{figure}[t]
    \centering
    \includegraphics[width = \linewidth]{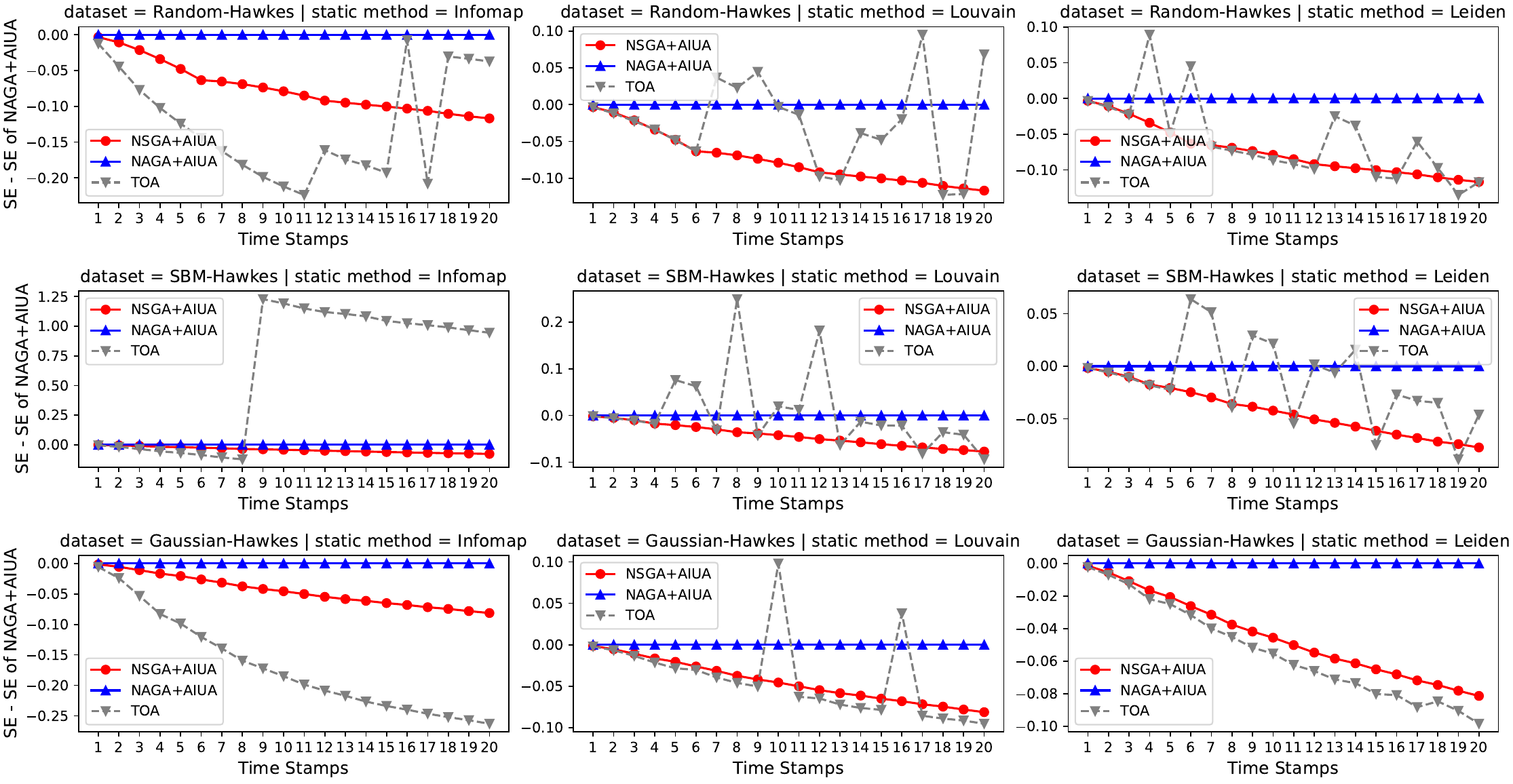}
    \caption{The updated structural entropy measured by \textit{NAGA+AIUA}, \textit{NSGA+AIUA}, and \textit{TOA} on artificial datasets with different static methods.
    Since the three curves for the artificial datasets are closer to each other than that for the real-world datasets, all displayed structural entropy values are subtracted from the structural entropy value of \textit{NAGA+AIUA} to better show the differences between the curves.
    }
    \label{fig:exp1ex}
\end{figure}

The experimental results are shown in Fig.~\ref{fig:exp1} \revised{(on real-world datasets) and Fig.~\ref{fig:exp1ex} (on artificial datasets)}.
Overall, the structural entropy obtained by \textit{NSGA+AIUA} based on the node-shifting strategy, is completely smaller than that obtained by \revised{\textit{NAGA+AIUA}} based on the naive adjustment strategy, in all settings.
For example, \revised{compared with \textit{NAGA+AIUA}}, \textit{NSGA+AIUA} reduces the updated structural entropy by up to about $12\%$ and $10\%$ on Facebook and DBLP respectively. 
This verifies that the node-shifting strategy is theoretically and practically able to reduce the structural entropy and represents that the strategy can obtain a significantly better encoding tree (community partitioning) than the naive adjustment strategy.

While maintaining high efficiency (evaluated in Section~\ref{sec:time}), our incremental algorithms still exhibit a performance that is close to or better than $TOA$.
\revised{In Fig.~\ref{fig:exp1}, the performance of \textit{NSGA+AIUA} is only slightly weaker than $TOA$ with Infomap on Cit-HepPh and Facebook.
Further, it even achieves lower structural entropy than the offline algorithm when Louvain and Leiden are chosen.
Two main reasons are listed as follows.
(1) In the incremental algorithms, the new community partitioning is obtained by locally modifying the original partitioning of the initial state.
While in $TOA$ the community partitioning is reconstructed globally from snapshots (updated graphs) by the static methods.
Although $TOA$ spends a large amount of time searching for optimal partitioning globally, there is no theoretical proof that it must be better than local and greedy optimization used in incremental algorithms.
(2) Additionally, the optimization objectives of incremental algorithms and $TOA$ are different.
The former focuses on the structural entropy itself while the latter aims to optimize modularity (Louvain and Leiden) or minimize the expected length of information transmission (Infomap).
}

\revised{
From the experimental results on artificial datasets (Fig.~\ref{fig:exp1ex}), we can conclude that our incremental algorithms have higher stability in the dynamic calculation of structural entropy.
In most cases, there are many mutations in the structural entropy value using \textit{TOA}, while our algorithms maintain a continuous and stable decrease.
It is because \textit{TOA} globally reconstructs the community for each iteration so that small node or edge changes may cause more influence.
By contrast, our algorithms dynamically adjust the community partitioning from the last snapshot, which affects only locally involved nodes.
}



\subsubsection{Hyperparameter Study}
In this part, we evaluate the influence on the updated structural entropy of different iteration numbers of the node-shifting adjustment strategy.
We use \textit{NSGA + AIUA} with iteration number $N = 3, 5, 7, 9$ to measure the mean updated structural entropy of the $20$ updated graphs, respectively, on each situation in Section~\ref{sec:app}.
As we can see from Table~\ref{tab:exp2}, the updated structural entropy decreases as the number of iterations increases most of the time.
The reason is that, as the number of iterations increases, more nodes will shift to their OPC, which leads to the further reduction of the structural entropy.
This experiment also demonstrates that our node-shifting adjustment strategy has excellent interpretability.

\begin{table}[t]
\fontsize{10}{10}\selectfont
\centering
\caption{The Updated Structural Entropy by Node-Shifting Adjustment Strategy with Different Number of Iterations.
\textbf{Bold} Number Denotes the Lowest Structural Entropy.}
\vspace{0.5em}
\begin{threeparttable}

\begin{tabular}{@{}cc|cccc@{}}
\toprule
\multicolumn{2}{c|}{$\#$ of iterations ($N$)}                        & $N = 3$       & $N = 5$      & $N = 7$                & $N = 9$                \\ \midrule
\multicolumn{1}{c|}{\multirow{3}{*}{Cit-HepPh}} & Infomap & 9.7155  & 9.7126  & 9.7122           & \textbf{9.7120}  \\
\multicolumn{1}{c|}{}                           & Louvain & 10.7804 & 10.7791 & 10.7786          & \textbf{10.7784} \\
\multicolumn{1}{c|}{}                           & Leiden  & 10.7802 & 10.7792  & 10.7788           & \textbf{10.7786}  \\ \midrule
\multicolumn{1}{c|}{\multirow{3}{*}{DBLP}} & Infomap & 7.3274  & 7.3255  & 7.3244           & \textbf{7.3242}  \\
\multicolumn{1}{c|}{}                           & Louvain & 9.4851  & 9.4842  & 9.4841           & \textbf{9.4838}  \\
\multicolumn{1}{c|}{}                           & Leiden  & 9.3919  & 9.3917  & 9.3912           & \textbf{9.3909}  \\ \midrule
\multicolumn{1}{c|}{\multirow{3}{*}{Facebook}}  & Infomap & 10.2134 & 10.2075 & 10.2060          & \textbf{10.2055} \\
\multicolumn{1}{c|}{}                           & Louvain & 11.2946 & 11.2898 & 11.2876          & \textbf{11.2864} \\
\multicolumn{1}{c|}{}                           & Leiden  & 11.3087 & 11.3052 & 11.3038          & \textbf{11.3030}          \\ \midrule
\multicolumn{1}{c|}{\multirow{3}{*}{Random-Hawkes}}    & Infomap & 11.7836\tnote{*} & 11.7832 & 11.7831          & \textbf{11.7831} \\
\multicolumn{1}{c|}{}                           & Louvain & 11.7836\tnote{*} & 11.7832 & 11.7831          & \textbf{11.7831} \\
\multicolumn{1}{c|}{}                           & Leiden  & 11.7836\tnote{*} & 11.7832 & 11.7831          & \textbf{11.7831}  
\\ \midrule
\multicolumn{1}{c|}{\multirow{3}{*}{Gaussian-Hawkes}}    & Infomap & 11.3352 & 11.3348 & 11.3347          & \textbf{11.3346} \\
\multicolumn{1}{c|}{}                           & Louvain & 11.3352 & 11.3348 & 11.3347          & \textbf{11.3346} \\
\multicolumn{1}{c|}{}                           & Leiden  & 11.3352 & 11.3348 & 11.3347          & \textbf{11.3346}  
\\ \midrule
\multicolumn{1}{c|}{\multirow{3}{*}{SBM-Hawkes}}    & Infomap & 11.6596 & 11.6595 & 11.6595          & \textbf{11.6595} \\
\multicolumn{1}{c|}{}                           & Louvain & 11.6596 & 11.6595 & 11.6595          & \textbf{11.6595} \\
\multicolumn{1}{c|}{}                           & Leiden  & 11.6596 & 11.6595 & 11.6595          & \textbf{11.6595} \\ 
\bottomrule
\end{tabular}
\begin{tablenotes}
    \item[*] \revised{
    The structural differences between different static methods on artificial Hawkes datasets are really small indicating that the initial community partitionings of the three methods are almost the same.
    }
\end{tablenotes}
\end{threeparttable}
\label{tab:exp2}
\end{table}

\subsubsection{Time Consumption Evaluation}\label{sec:time}
\begin{figure}[!t]
    \centering
    \includegraphics[width = 0.85\linewidth]{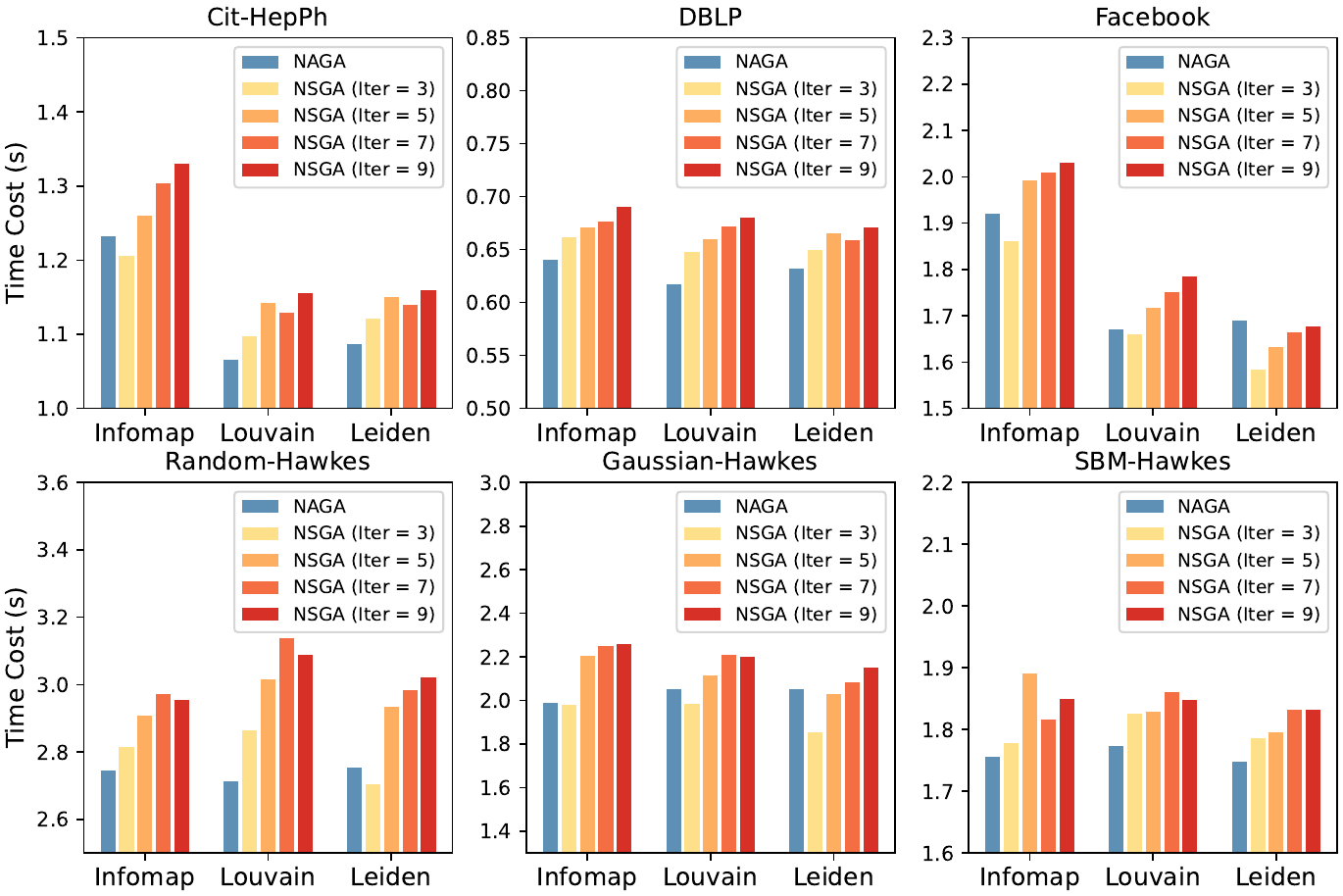}
    \caption{Mean time consumption of \textit{NAGA+AIUA} and \textit{NSGA+AIUA} ($N = 3, 5,7,9$) over $20$ time stamps on each dataset under different static methods.}
    \label{fig:exp3}
\end{figure}

Fig.~\ref{fig:exp3} shows the time consumption comparison between our incremental algorithms, i.e. \textit{NAGA+AIUA} and \textit{NSGA+AIUA} ($N = 3, 5,7,9$), on all $6$ datasets.
The vertical axis in the figure represents the mean time consumption of the chosen incremental algorithm across all $20$ snapshots.
The horizontal axis represents $3$ selected static methods.
As we can see, the time cost of \textit{NSGA+AIUA} increases as the increase of iteration number $N$.
In addition, the time consumption of \textit{NAGA+AIUA} is less than \textit{NSGA+AIUA} with $N = 5$ in most cases.

\revised{
Table~\ref{tab:time} shows the time comparison between our online algorithm \textit{NSGA+} \textit{AIUA} ($N=5$) and the offline algorithm \textit{TOA}.
As we can see from the results, all our proposed incremental algorithms are significantly faster than the existing static methods.
Specifically, for example, \textit{NSGA+AIUA} ($N=5$) obtain over $140.93$x and $77.81$x speed up on average on DBLP and SBM-Hawkes, respectively, in contrast with the static method using Infomap.
}

\begin{table}[!h]
\fontsize{10}{9}\selectfont
\centering
\caption{\revised{Time Consumption Comparison of Our Incremental Algorithms (Online Time) and the Baseline Traditional Offline Algorithm (Offline Time).}}

\resizebox{\columnwidth}{!}{\begin{threeparttable}

\vspace{0.5em}
\begin{tabular}{@{}l|ccc|ccc|ccc@{}}
\toprule
Dataset       & \multicolumn{3}{c|}{Cit-HepPh}                   & \multicolumn{3}{c|}{DBLP}            & \multicolumn{3}{c}{Facebook}                     \\ \midrule
Static Method & \multicolumn{1}{l}{Infomap}  & Louvain & Leiden  & Infomap     & Louvain   & Leiden     & \multicolumn{1}{l}{Infomap}  & Louvain & Leiden  \\ \midrule
Online Time\tnote{1} &
  \multicolumn{1}{c}{$1.23s$\tnote{2}} &
  $1.07s$ &
  $1.09s$ &
  $0.64s$ &
  $0.62s$ &
  $0.63s$ &
  \multicolumn{1}{c}{$1.92s$} &
  $1.67s$ &
  $1.69s$ \\
Offline Time  & \multicolumn{1}{c}{$49.96s$} & $6.42s$ & $5.05s$ & $49.80s$    & $4.72s$   & $12.30s$   & \multicolumn{1}{c}{$80.76s$} & $8.15s$ & $7.43s$ \\ \midrule
Speedup\tnote{3} &
  \multicolumn{1}{l}{$\uparrow40.62$x} &
  $\uparrow6.00$x &
  $\uparrow4.63$x &
  $\uparrow77.81$x &
  $\uparrow7.61$x &
  $\uparrow19.52$x &
  \multicolumn{1}{l}{$\uparrow42.06$x} &
  $\uparrow4.88$x &
  $\uparrow4.40$x \\ 
  \bottomrule
  \toprule
Dataset       & \multicolumn{3}{c|}{Random-Hawkes}               & \multicolumn{3}{c|}{Gaussian-Hawkes} & \multicolumn{3}{c}{SBM-Hawkes}                   \\ \midrule
Static Method & Infomap                      & Louvain & Leiden  & Infomap     & Louvain   & Leiden     & Infomap                      & Louvain & Leiden  \\ \midrule
Online Time   & $2.21s$                      & $2.48s$ & $1.32s$ & $1.48s$     & $1.52s$   & $1.55s$    & $1.76s$                      & $1.55s$ & $1.05s$ \\
Offline Time  & $58.37s$                     & $6.06s$ & $3.02s$ & $18.72s$    & $4.08s$   & $3.69s$    & $248.04s$                     & $4.01s$ & $3.02s$ \\ \midrule
Speedup &
  $\uparrow26.41$x &
  $\uparrow2.44$x &
  $\uparrow2.28$x &
  $\uparrow12.65$x &
  $\uparrow2.68$x &
  $\uparrow2.38$x &
  $\uparrow140.93$x &
  $\uparrow2.59$x &
  $\uparrow2.88$x \\ \bottomrule
\end{tabular}%

\begin{tablenotes}
    \item[1] Time cost of \textit{NSGA+AIUA} ($N=5$).
    \item[2] Mean time consumption over $20$ time stamps.
    \item[3] Speedup = Offline Time/Online Time.
    
\end{tablenotes}
\label{tab:time}
\end{threeparttable}}

\end{table}

\newrevised{
The reason why we choose $N = 5$ as the default parameter is as follows.
As shown in Fig.~\ref{fig:exp3}, the time consumption of the node-shifting strategy rises linearly with $N$. 
However, Table~\ref{tab:exp2} shows that the rate of decline of structural entropy gradually decreases from $N = 3$ to $N = 9$, and the structural entropy values of $N = 7$ and $N = 9$ are very close.
That is, the optimization efficiency of structural entropy decreases with increasing $N$. 
To seek a balance between efficiency and effectiveness, we take the compromise value $N = 5$.
}

\subsubsection{Update Threshold Analysis}

In scenarios with minimal changes, updating structural information might not be necessary. 
In this part, we set a threshold for the magnitude of graph changes before initiating updates to cut total time consumption.
Specifically, the updated structural entropy will not be calculated until the incremental edge number exceeds a certain percentage (referred to as the update threshold $\theta$) of the edge number of the last updated graph.
As we can see from Table~\ref{tab:thresh}, the total time reduces $63\%$-$87\%$ with \textit{NAGA+AIUA} and $47\%$-$72\%$ with \textit{NSGA+AIUA} when $\theta$ is set from $5\%$ to $20\%$.
Meanwhile, the fluctuation of the final structural entropy remains within $0.15\%$, which indicates that the threshold has little impact on the precision of structural entropy.
Overall, the setting of the updated threshold leads to improved efficiency and better adaptation to graphs undergoing frequent alterations.

\begin{table}[!h]
\centering
\caption{
\revised{The Influence on Total Time Consumption and Structural Entropy of Different Updated Threshold $\theta$ on Cit-HepPh dataset with Infomap static method.}
}
\vspace{0.3em}
\resizebox{\columnwidth}{!}{%
\begin{threeparttable}
\fontsize{9}{10}\selectfont
\begin{tabular}{@{}cc|ccccc@{}}
\toprule
\multicolumn{2}{c|}{Updated Threshold}                      & $\theta = 0\%$ & $\theta = 5\%$              & $\theta = 10\%$   & $\theta = 15\%$   & $\theta = 20\%$   \\ \midrule
\multicolumn{1}{c|}{\multirow{2}{*}{NA\tnote{1}}} & Total Time & $32.89s$       & $12.20s$\,\scriptsize$(\downarrow 63\%)$ \tnote{2} & $9.16s$\,\scriptsize($\downarrow 72\%$) & $5.64s$\,\scriptsize($\downarrow 83\%$) & $4.44s$\,\scriptsize($\downarrow 87\%$) \\
\multicolumn{1}{c|}{} & Final SE & $10.9736$ & $10.9687$\,\scriptsize$(\le 0.1\%)$ \tnote{3}& $10.9690$\,\scriptsize$(\le 0.1\%)$ & $10.9668$\,\scriptsize$(\le 0.1\%)$ & $10.9681$\,\scriptsize$(\le 0.1\%)$ \\ \midrule
\multicolumn{1}{c|}{\multirow{2}{*}{NS\tnote{1}}} & Total Time & $24.83s$       & $13.11s$\,\scriptsize$(\downarrow 47\%)$           & $10.52s$\,\scriptsize$(\downarrow 58\%)$ & $8.27s$\,\scriptsize$(\downarrow 67\%)$ & $6.93s$\,\scriptsize$(\downarrow 72\%)$ \\
\multicolumn{1}{c|}{} & Final SE & $10.0269$ & $10.0243$\,\scriptsize$(\le 0.1\%)$ & $10.0186$\,\scriptsize$(\le 0.1\%)$ & $10.0352$\,\scriptsize$(\le 0.1\%)$ & $10.0419$\,\scriptsize$(\le 0.15\%)$ \\ \bottomrule
\end{tabular}

\begin{tablenotes}
    \scriptsize
    \item[1] ``NA/NS'' represents \textit{NAGA/NSGA+AIUA}.
    \item[2] The percentage reduction in time consumption compared to the case where $\theta = 0\%$.
    \item[3] The percentage fluctuation of the final structural entropy compared to the case where $\theta = 0\%$.
    
\end{tablenotes}
\label{tab:thresh}
\end{threeparttable}
}

\end{table}

\revised{\subsubsection{Robustness Analysis}}

\begin{figure}[!h]
    \centering
    \includegraphics[width = \linewidth]{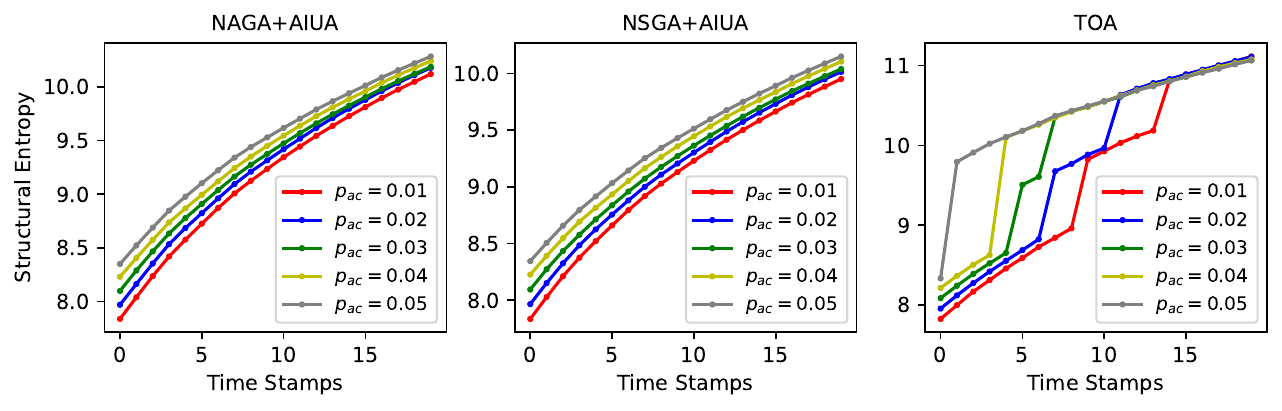}
    \caption{
    \revised{The structural entropy of \textit{NAGA/NSGA+AIUA} and \textit{TOA} with Infomap under different initial state parameter $p_{\text{ac}}$ representing different noise conditions.}
    }
    \label{fig:noise}
\end{figure}

In this subsection, we evaluate the robustness of the proposed framework on $5$ new artificial datasets with different noise conditions.
Specifically, we generate $5$ initial states using the random partition method mentioned in Section~\ref{sec:ad} with $p_\text{ac}$ (representing the noise) ranging from $0.01$ to $0.05$ and $\mathcal{S} = [100, 100, 200, 200, 200]$.
For the convenience of comparison, the total edge number of the initial state is kept nearly the same, fluctuating between $37357$ and $38175$, by manually selecting the appropriate $p_\text{in}$.
After the generation of the $5$ initial states, we use the Hawkes Process to generate $74714$ incremental edges, around $2$ times the initial edges, and split them into $20$ sub-sequences as the incremental of $20$ time stamps.
Then we use \textit{NAGA/NSGA+AIUA} and \textit{TOA} to calculate the structural entropy for all $20$ updated graphs and the results are shown in Fig.~\ref{fig:noise}.

As we can see from Fig.~\ref{fig:noise}, the structural entropy curves using \textit{NAGA/ NSGA+AIUA} are smooth no matter what $p_\text{ac}$ takes.
However, one or two jumps in structural entropy take place when it comes to $TOA$.
Furthermore, the higher the noise ($p_\text{ac}$) is, the earlier and larger the structural entropy mutates, and the more unstable the $TOA$ algorithm is.
Additionally, upon examination, the number of communities does not change, meaning that these abrupt changes do not stem from changes in the number of communities, but due to dramatic changes in their content.
Therefore, we can conclude that our algorithm maintains the structural entropy at a stable and lower level due to the property of keeping the original community structure from changing drastically, showing high robustness to the increasing noise.

In addition, we note that the structural entropy curve gradually shifts upward as $p_\text{ac}$ increases. 
This is because the numbers of edges of the initial states have slight errors.
Specifically, the larger the $p_\text{ac}$, the more initial edges there are (from $37357$ to $38175$).

\subsubsection{\newrevised{Gap between Incre-2dSE and the Current Static Structural Entropy Measurement Method}}

\begin{figure}
    \centering
    \includegraphics[width = 0.8\linewidth]{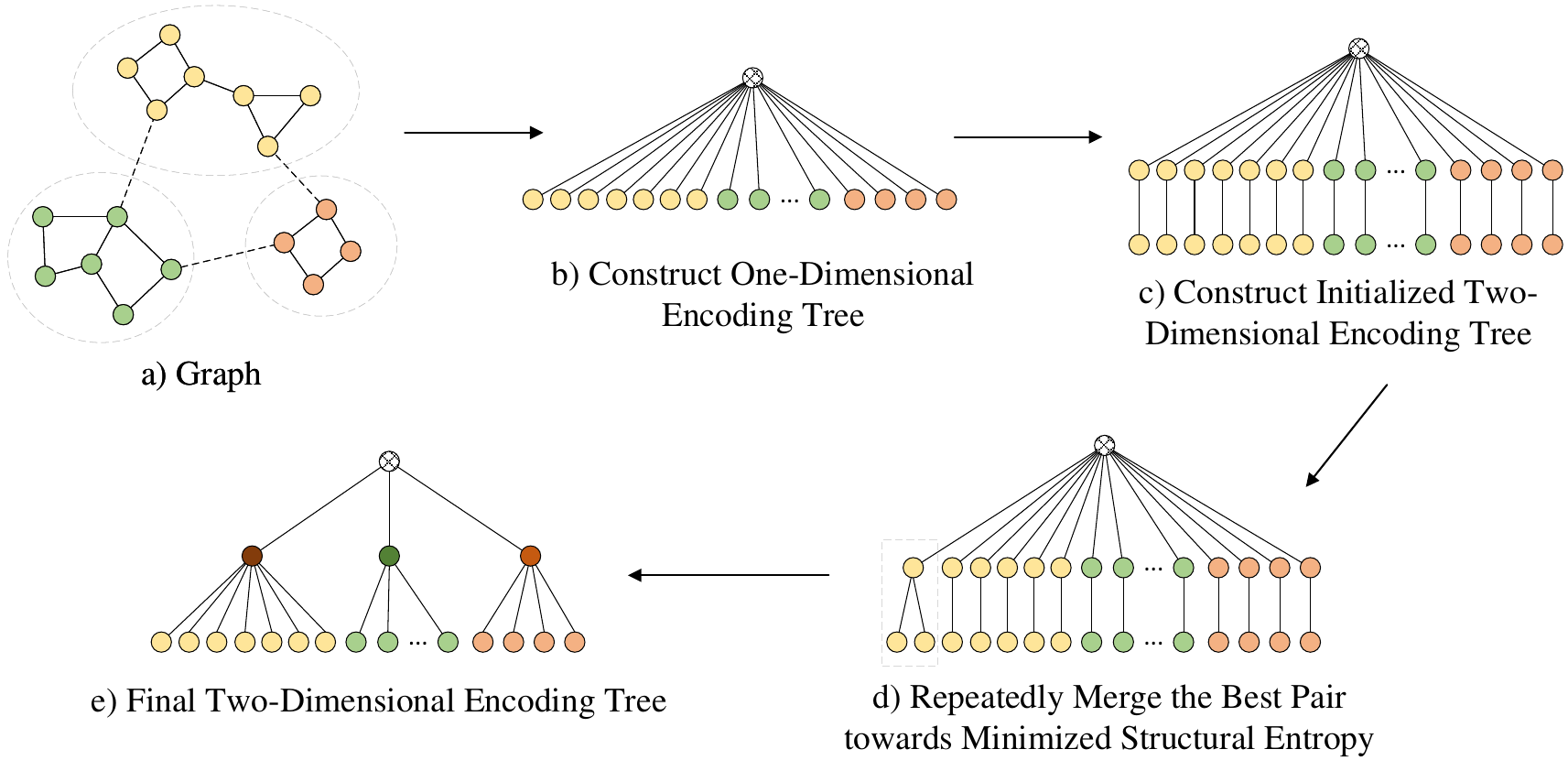}
    \caption{\newrevised{An illustration of two-dimensional structural entropy minimization process.}}
    \label{fig:2d-SEM}
\end{figure}

\newrevised{
In this part, we try to study the gap between \textit{Incre-2dSE} and the current static algorithms.
The current mainstream static algorithm for structural entropy measurement is named \textit{structural entropy minimization \textit{(SEM)}}~\citep{structural-entropy}, which is a greedy $k$-dimensional encoding tree construction algorithm for static graphs whose objective function is structural entropy. 
Fig.~\ref{fig:2d-SEM} gives an illustration of the \textit{SEM} algorithm in two-dimensional cases (\textit{2d-SEM}). 
Fig.~\ref{fig:2d-SEM}(a) is an example graph. 
We first construct a one-dimensional encoding tree for this graph (Fig.~\ref{fig:2d-SEM}(b)). 
The one-dimensional encoding tree has one root and $|\mathcal{V}|$ leaf nodes. 
Next, we add a successor node to each leaf node to construct an initialized two-dimensional encoding tree (Fig.~\ref{fig:2d-SEM}(c)). 
Then we select the best pair of $1$-height nodes to merge them which minimizes the structural entropy (Fig.~\ref{fig:2d-SEM}(d)). 
At last, We repeat this merging step until the structural entropy doesn’t go down to get the final two-dimensional encoding tree (Fig.~\ref{fig:2d-SEM}(e)).
}

\begin{figure}
    \centering
    \includegraphics[width = \linewidth]{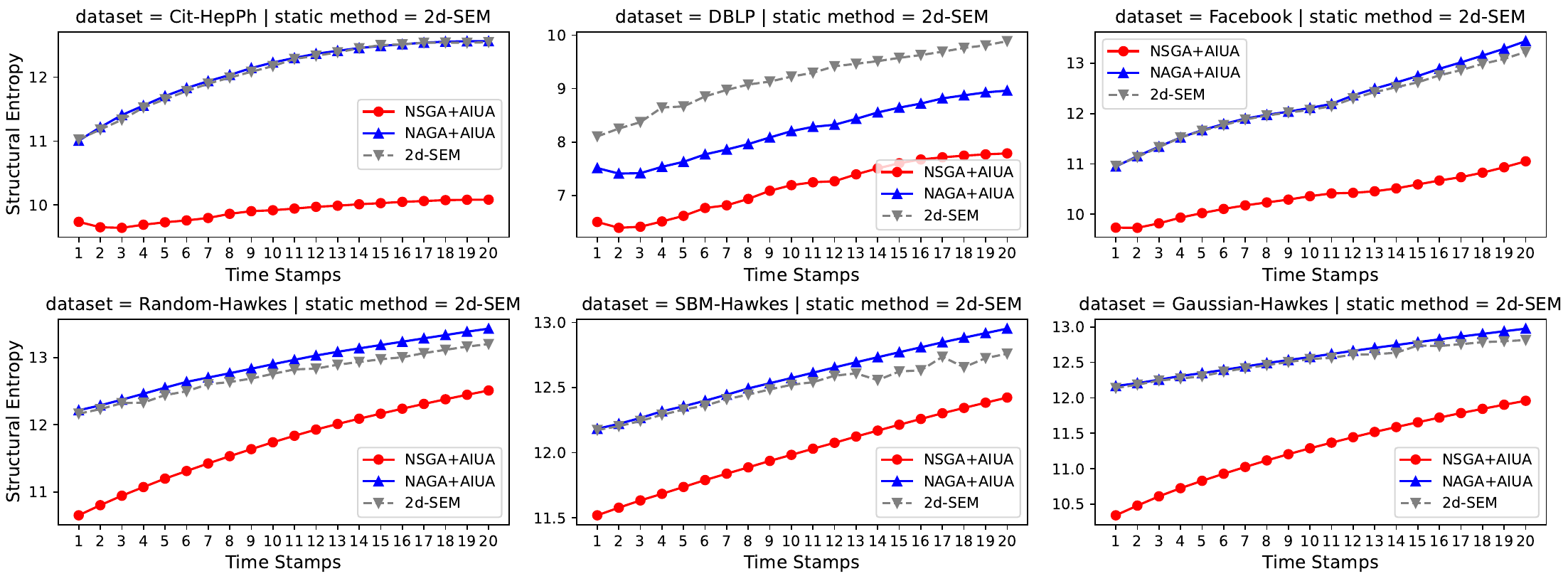}
    \caption{\newrevised{An illustration of two-dimensional structural entropy minimization process.}}
    \label{fig:2d-SEM-exp}
\end{figure}

\newrevised{
We measure the structural entropy of \textit{Incre-2dSE} (\textit{NAGA/NSGA+AIUA}) and \textit{2d-SEM}\footnote{\newrevised{https://github.com/RingBDStack/SITool/tree/main/python}} through all timestamps like Section~\ref{sec:app} on the six datasets mentioned in Table~\ref{tab:datasets}. 
According to our experimental results (Fig.~\ref{fig:2d-SEM-exp}), there is a gap between the structural entropy of \textit{2d-SEM} and our dynamic framework \textit{Incre-2dSE} in some cases. 
Specifically, on all datasets except DBLP, the structural entropy of \textit{NAGA+AIUA} is higher than but very close to \textit{2d-SEM}. 
On DBLP, \textit{NAGA+AIUA} is better than \textit{2d-SEM}. 
On the contrary, \textit{NSGA+AIUA} performs significantly better than \textit{2d-SEM} on all datasets. 
In addition, the gap between \textit{NSGA+AIUA} and \textit{2d-SEM} is gradually increasing on Cit-HepPh and Facebook, is gradually decreasing on Random-Hawkes and Gaussian-Hawkes, and remains almost unchanged on DBLP and SBM-Hawkes.
The reason why there is a gap between \textit{Incre-2dSE} and \textit{2d-SEM} is the same as the reason why there is a gap between \textit{Incre-2dSE} and \textit{TOA} in Section~\ref{sec:app}—\textit{Incre-2dSE} updates the community and encoding tree locally and incrementally while \textit{2d-SEM} constructs encoding trees globally from scratch for each snapshot.
}

\subsubsection{Convergence Evaluation}
\begin{figure}[ht]
\centering
\includegraphics[scale=0.4]{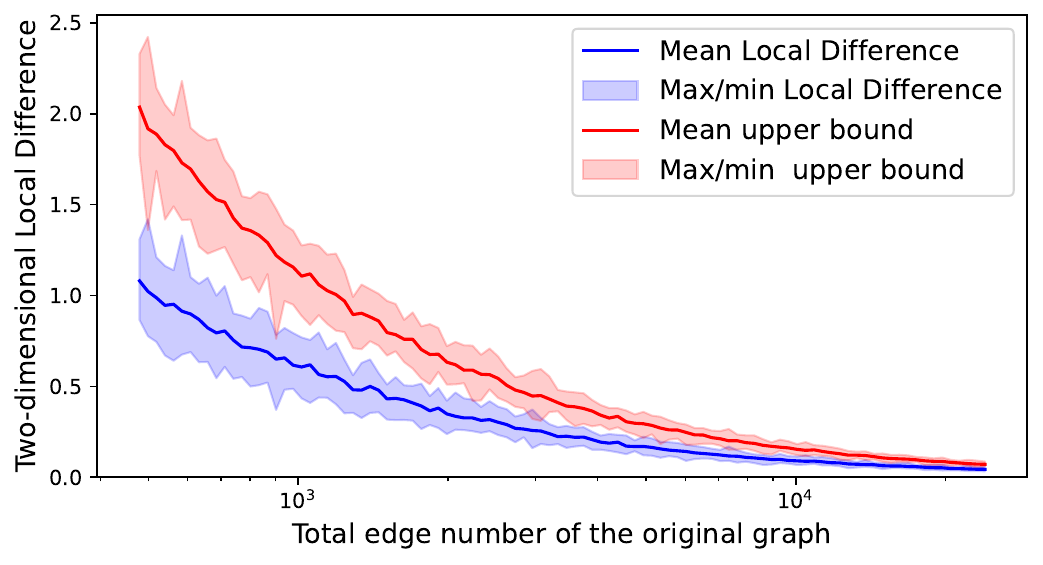}
\caption{The statistics of the Local Difference and its upper bound.}
\label{moment_plot_figure_2d}
\end{figure}

In this part, we conduct a statistical experiment to confirm the convergence of the Local Difference and its first-order absolute moment.
We first generate artificial original graphs with increasing total edge numbers from $480$ to $24000$.
Based on each original graph, we generate $30$ incremental sequences with size $n$ sampling from a normal distribution with a mean of $100$ and a standard deviation of $10$.
These incremental sequences are generated by repeatedly adding edges within a community with probability $p_\mathrm{pin}\in [0,0.8)$, adding edges across two random communities with probability $p_\mathrm{pac}\in[0,0.1)$, and adding nodes with probability $p_\mathrm{n} = 1 - p_{pin} - p_{pac}$.
We then count the Local Difference and its upper bound.
The results are shown in Fig.~\ref{moment_plot_figure_2d}.
As the total edge number increases from $1$ to $50$ times, the mean Local Difference gradually decreases by $95.98\%$ (from $1.08$ to $0.04$), respectively, and is always positive.
This solidly supports the convergence of the Local Difference and its first-order absolute moment.
Moreover, the Local Difference is always below its upper bound, which confirms the validity of our bound.

\subsubsection{\revised{One-Dimensional Structural Entropy Measurement of Directed Weighted Graphs}}

In this subsection, we evaluate the time consumption of the two approximate one-dimensional structural entropy measurement methods, Global Aggregation and Local Propagation, on two artificial datasets, i.e., Erdős-Rényi (ER)~\citep{erdds1959random} and Cycle.

\noindent \textbf{\revised{ER initial state.}}
\revised{
The initial state of the ER dataset is created by the Erdős-Rényi model implemented by \textit{erdos\_renyi\_graph} in Networkx~\citep{networkx} which has two main parameters.
One is the number of nodes $n$ and the other is the probability of edge creation $p$.
In this dataset, the Erdős-Rényi model chooses each of the $n(n-1)$ possible directed edges with probability $p$.
}

\noindent \textbf{\revised{Cycle initial state.}}
\revised{
The Cycle dataset's initial state contains cyclically connected nodes, where the edge direction is in increasing order.
For example, a cycle dataset $\{v_0, v_1, ..., v_n\}$ has $n+1$ direct edges $\{(v_0, v_1), (v_1, v_2), ..., (v_{n-1}, v_n)$ $, (v_n, v_0)\}$.
}

\noindent \textbf{\revised{Weight and incremental settings.}}
\revised{
For the initial states of ER and Cycle datasets, we first set all edge weights as random integers in $[1,10]$.
Then we generate several incremental sequences with sizes $500$, $2000$, $5000$, and $10000$, respectively, corresponding to four different updated graphs from each initial state.
Specifically, for each incremental edge, we randomly choose two nodes to add a direct edge between them with random weight in $[1,10]$.
If the edge exists, we randomly change its weight as another value in $[1,10]$.
}

\noindent \textbf{\revised{Experimental settings.}}
For each initial state, we use Global Aggregation with iteration number $50$ to calculate the initial stationary distribution from a uniform distribution.
For each initial state, we then use Global Aggregation and Local Propagation to iteratively calculate the structural entropy of the $4$ updated graphs and record the time consumption of each iteration.

\begin{figure}[!h]
    \centering
    \includegraphics[width = \linewidth]{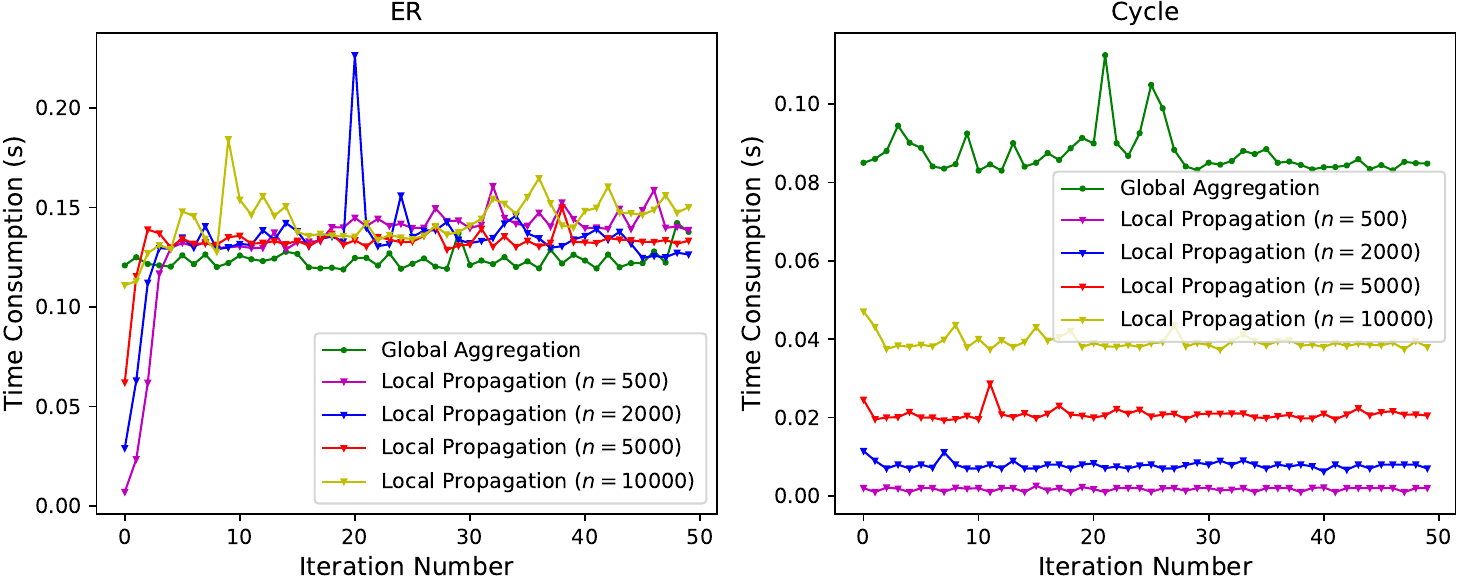}
    \caption{
    \revised{
    Time consumption of Global Aggregation and Local Propagation on ER and Cycle dataset.
    Note that only the time consumption of Global Aggregation of $n=500$ is shown since the theoretical time consumption of different incremental size $n$ is the same.
    }
    }
    \label{fig:exp1d}
\end{figure}

\noindent \textbf{\revised{Experimental results.}}
The time consumption experimental results are shown in Fig.~\ref{fig:exp1d}.
On the ER dataset, Local Propagation is faster than Global Aggregation only in the first several iterations.
Besides, the time consumption is lower when $n$ is smaller.
This is because the time consumption of each iteration of Local Propagation is approximately linear with the size of the involved node set.
The smaller the incremental size $n$, the number of the involved nodes is less.
However, since the mean number of the successors of ER graphs is more than $1$, the involved nodes will rapidly expand into the entire node set.
Therefore, when the iteration number is larger than $5$, the efficiency of Local Propagation is nearly the same as that of Global Aggregation.
On the contrary, in the Cycle dataset, the mean number of the direct successors of each node is nearly $1$.
That is, the size of the involved node set will not change, so the time consumption of Local Propagation will not increase as the iteration number grows.
In addition, obviously the larger the incremental size $n$, the higher the time consumption whatever the iteration number is.

\section{Related Works}\label{sec:relatedworks}
\noindent \textbf{Graph Entropy.}
Many efforts have been devoted to measuring the information in graphs. 
The graph entropy was first defined and investigated by Rashevsky~\citep{rashevsky1955life-c1}, Trucco~\citep{trucco1956note-c2}, and Mowshowitz~\citep{mowshowitz1967entropy}.
After that, a different graph entropy closely related to information and coding theory was proposed by Körner\citep{korner1973coding-c7}.
Later, Bianconi~\citep{bianconi2009entropy} introduced the concept of ``structural entropy of network ensembles", known as the Gibbs entropy.
Anand et al.~\citep{anand2009entropy} proposed the Shannon entropy for network ensembles. 
Braunstein et al.~\citep{braunstein2006laplacian} proposed the von Neumann graph entropy based on the combinatorial graph Laplacian matrix.
These three metrics \citep{bianconi2009entropy, anand2009entropy, braunstein2006laplacian} are defined by statistical mechanics and are used to compare different graph models.
However, most of the current graph entropy measures are based on the Shannon entropy definition for probability distributions, which has significant limitations when applied to graphs~\citep{zenil2017low}.
Recently, many efforts have been devoted to capturing the dynamic changes of graphs, e.g., the research based on the Algorithmic Information Theory~\citep{zenil2019causal, zenil2019algorithmic}.
The structural entropy method used in this paper proposed by Li et al.~\citep{structural-entropy} provides an approach to measuring the high-dimensional information embedded in graphs and can further decode the graph's hierarchical abstracting by an optimal encoding tree.
This method is widely used in the fields of graph learning~\citep{zou2023se}, reinforcement learning~\citep{zeng2023effective}, and social networks~\citep{cao2024hierarchical}.

\noindent \textbf{Fast Computation for Graph Entropy.}
Chen et al.~\citep{chen2019fast} proposed a fast incremental von Neumann graph entropy computational framework, which reduces the cubic complexity to linear complexity in the number of nodes and edges.
Liu et al.~\citep{liu2022similarity, liu2021bridging} used the structural entropy~\citep{structural-entropy} as a proxy of von Neumann graph entropy for the latter's fast computation and also implemented an incremental method for one-dimensional structural entropy \revised{on undirected graphs}.
In this paper, we mainly focus on the incremental computation for two-dimensional structural entropy based on our dynamic adjustment strategies for encoding trees.
\revised{Besides, we also discuss the computation method for one-dimensional structural entropy on directed weighted graphs.}

\section{Conclusion} \label{sec:conc}
In this paper, we propose two novel dynamic adjustment strategies, namely the naive adjustment strategy and the node-shifting adjustment strategy, to analyze the updated structural entropy and incrementally adjust the original community partitioning towards a lower structural entropy.
We also implement an incremental framework, i.e., supporting the real-time measurement of the updated two-dimensional structural entropy.
\revised{Further, we discuss the extension of our proposed methods to weighted graphs and the one-dimensional structural entropy computation on directed and weighted graphs.}
In the future, we aim to develop more dynamic adjustment strategies for hierarchical community partitioning and incremental measurement algorithms for higher dimensional structural entropy.


\section*{Acknowledgments}
This work is supported by the National Key R\&D Program of China through grant 2022YFB3104703, NSFC through grants 62322202 and 61932002, Beijing Natural Science Foundation through grant 4222030, Guangdong Basic and Applied Basic Research Foundation through grant 2023B1515120020, CCF-DiDi GAIA Collaborative Research Funds for Young Scholars, and the Fundamental Research Funds for the Central Universities.


\bibliographystyle{elsarticle-num} 
\bibliography{ref}






\end{document}